\documentclass[preprintnumbers,amsmath,amssymb,nofootinbib
,superscriptaddress]{revtex4}

\usepackage{hyperref}
\usepackage{graphicx}
\usepackage{color}
\usepackage{xcolor}
\usepackage{comment}
\usepackage{lmodern}
\usepackage{amssymb}
\usepackage{lipsum}
\usepackage{mwe}

\newcommand{\backo}{\!\!\!\!\!\!\!\!\!\!}
\newcommand{\ebo}{\eea }
\newcommand{\bbo}{\bea  && }
\newcommand{\smebo}{\eea  }
\newcommand{\smbbo}{\bea  && }

\newcommand{\bl}{\biggl(}
\newcommand{\br}{\biggr)}
\newcommand{\Tr}{\makebox{Tr}}

\newcommand{\vvr}{\vec{r}}

\newcommand{\vvq}{\vec{q}}

\newcommand{\be}{\begin{equation}}  
\newcommand{\ee}{\end{equation}}  
\newcommand{\bea}{\begin{eqnarray}}   
\newcommand{\eea}{\end{eqnarray}}  
\newcommand{\ba}{\begin{array}}  
\newcommand{\ea}{\end{array}}

\usepackage{diagbox}

\newskip\humongous \humongous=0pt plus 1000pt minus 1000pt

\newif\ifdtup

\def\oldreffmt#1{\rlap{[#1]} \hbox to 2\parindent{}}

\def\figfmt#1{\rlap{Figure {#1}} \hbox to 1in{}}  
  
\def\Tr{\mathop{\rm Tr}}

\def\slash#1{#1\!\!\!\!/\!\,\,} 	
\def\beq{\begin{equation}}  
\def\eeq{\end{equation}}  
\def\bea{\begin{eqnarray}}  
\def\eea{\end{eqnarray}}  
\def\half{\frac{1}{2}}  
  
\def\bq{\begin{quote}}  
\def\eq{\end{quote}}

\def\half{\frac{1}{2}}    
\newcommand{\nonumbo}{ \nonumber \\ && }
\def \lta {\mathrel{\vcenter  
     {\hbox{$<$}\nointerlineskip\hbox{$\sim$}}}}  
\def \gta {\mathrel{\vcenter  
     {\hbox{$>$}\nointerlineskip\hbox{$\sim$}}}}   

\relax

\newdimen\tdim  
\tdim=\unitlength  
\def\bar{\overline}

\RequirePackage{caption}
\begin{document}

\preprint{FERMILAB-PUB-25-0219-T}

\title{Natural Top Quark Condensation\\ (a Redux)
}
\author{Christopher T. Hill}
\email{chill35@wisc.edu}
\affiliation{Fermi National Accelerator Laboratory,
P. O. Box 500, Batavia, IL 60510, USA}
\affiliation{Department of Physics, University of Wisconsin-Madison, Madison, WI, 53706
}

\begin{abstract}
 
The Nambu--Jona-Lasinio (NJL) model involves a pointlike 4-fermion interaction.  While it gives 
a useful description of chiral dynamics (mainly in QCD), it
nonetheless omits the crucially important internal wave-function of a two-body bound state, $\phi(r)$.
This becomes significant near critical coupling where $\phi(r)$ extends to large
distance,
leading to dilution and suppression of induced couplings $\propto \phi(0)$, such as the Yukawa and quartic couplings, as well as reduced fine-tuning of a hierarchy.  In top quark condensation, where 
the Brout-Englert-Higgs (BEH) boson is a $\bar{t}t$ bound state and we have a UV completion such as topcolor, we must go beyond the NJL model and include
effects of $\phi(r)$.  We provide a formulation of this for the BEH boson, and find
that it leads to an extended $\phi(r)$,  a significantly reduced and natural composite scale of $M_0 \sim 6 $  TeV,
a successful prediction for the quartic coupling, $\lambda$,
and fine--tuning that is reduced to a few percent, providing
 a compelling candidate solution to the naturalness problem of
the BEH boson.  The theory is testable and the associated new physics 
may soon emerge at LHC energy scales.

\end{abstract}

\maketitle
 
\date{\today}


\email{chill35@wisc.edu}


\section{Introduction }

In the early 1990's we proposed the idea of ``top quark condensation,'' 
i.e., that the  Brout-Englert-Higgs (BEH) boson is composed of top + anti-top quarks
\cite{Nambu1}$-$\cite{NSD}.
The minimal  model introduced a 4-fermion pointlike
interaction, at large mass scale, $M_0$, amongst third generation
quarks, 
 \bbo
 \label{oneone}
\frac{g_0^2}{M_0^2}
[\bar{\psi}_{iL}(x)\psi_{R}(x)]\;[\bar{\psi}_R(x)\psi^i_{L}(x)]; \qquad \;\;\; \psi^i_L= \bl\begin{array}{c} t\\ b \end{array}\br_L, \;\;\;\psi_R = t_R,
\ebo
(where $i$ is an electroweak $SU(2)$ index, and  $[...]=$ denotes a sum over color).
To treat this we deployed the Nambu-Jona-Lasinio (NJL) model \cite{NJL}, and
introduced significant renormalization group (RG) improvement \cite{BHL}.
The NJL model
led to a composite BEH electroweak isodoublet, 
described by a local field, $H^i(x)\sim[\bar{\psi}_R(x)\psi^i_L(x) ]$.
When tuned to the known electroweak scale, $v_{weak}=175$ GeV, the theory predicted
the top quark and BEH boson masses.

The minimal top-condensation theory was one of the earliest composite BEH models,
and was ``philosophically
 successful'' in that it non-trivially tied together
unrelated parameters of the Standard Model (SM).   
However, the explicit predictions of the model
were ultimately ruled out by subsequent experiment: $m_{top}
\approx 220$ GeV, (cf. $175$ GeV, experiment) and  $m_{BEH} \approx 260$ GeV (cf. $125$ GeV, experiment).  The
predicted $m_{BEH}$ corresponded to a SM quartic coupling prediction of $\lambda \approx 1$, (cf.
$\lambda \approx 0.25$, experiment), 
while the predicted $m_{top}$ corresponded to a SM Yukawa coupling prediction of $g_{top}\equiv g_Y \approx 1.25$, (cf.
$g_Y \approx 1.00$, experiment).

Moreover,
to accommodate the electroweak scale, the theory required
 ultra-large $M_0\sim 10^{15}$ GeV, which implied a drastic fine--tuning of the coupling $g_0^2$ to its critical value $g_c^2$, 
 \bbo
\bl 1-\frac{g_0^2}{g_c^2}\br \sim \frac{v_{weak}^2}{M_0^2}\sim 10^{-26}\;(!)
\ebo 
So,  the top condensation theory was directly testable by experiment and it evidently
failed.

The NJL model was used in top condensation since it is concise, manifestly Lorentz invariant and provides a guide to the chiral
symmetry breaking dynamics seen in QCD.  There it leads to useful results, where the 
fundamental chiral current quarks dynamically become the heavy constituent quarks, 
yielding the chiral--constituent quark model \cite{Manohar}.
The NJL model builds the pseudoscalar mesons and $\sigma$ field by
``integrating out'' the light quarks which, in a sense, imitates quark
confinement. It does well at explaining the Gasser-Leutwyler coefficients and other parameters \cite{Bijnens}\cite{NJLReview} and it remarkably predicted
a universal ``chiral mass
gap'' in all heavy-light quark bound states, leading to long lived heavy-strange
resonances, such as the  $D_s(2317)$, \cite{bardeenhill}.

However,  in the case of
top condensation  one has a bound state of non-confined constituents.
Hence the bound state couples to free unbound fermions that have
the same quantum numbers as the constituents.
Here we encounter fundamental physical limitations of the NJL model:
\begin{itemize}
\item the NJL model is an effective pointlike 4-fermion interaction
associated with a ``large'' mass scale $M_0$, and the resulting bound states emerge as {\em pointlike} fields with mass $\mu^2< M_0^2$;
\item  in the NJL model 
the binding mechanism is entirely driven by quantum loop effects, while 
we see in nature that binding readily occurs
semiclassically without quantum loops, such as the hydrogen atom;
\item Mainly, the NJL model lacks an internal wave-function $\phi(r)$. The
inclusion of $\phi(r)$ has significant impact upon the conclusions drawn from the model.
\end{itemize}

In the case of the hydrogen atom, before turning on the Coulomb interaction, there
are open scattering states involving free protons and electrons.\footnote{We refer to compact bound state wave-functions as ``normalizable.'' Open 
free particle scattering states, that require typically ``box normalization,'' are
referred to as ``non-normalizable,'' requiring introduction of some IR cut-off on the wave-functions.}  As the interaction
is turned on the lowest energy scattering states flow to become the bound states, 
while most scattering states remain unbound. 
The dynamics is governed by the non-relativistic semiclassical (tree level)
Schr\"odinger equation \cite{Schrodinger}, leading
to normalizable yet spatially extended  wave-functions on the scale $(\alpha m_e)^{-1}$.
The atom is described naturally in a configuration space picture.
Quantum loop effects (such as the Lamb shift) are higher order corrections to this mostly semiclassical phenomenon.

In the NJL model the picture is substantially different. There is no semiclassical binding
producing an extended bound state. Rather, the bound state is described by a local effective field, $\Phi(x)$, with its properties arising from quantum loops.  The loops integrate out the constituent
fermions from the large mass scale of
the interaction, $M_0$, down to an IR cut-off $\mu$
(e.g., $M_0\sim 1$ GeV and $\mu \sim f_\pi \sim 100$ MeV in QCD). When the discussion is formulated
in momentum space, treated in the
large $N_{color}$ limit,  bound states appear as poles in the S-matrix upon summing
 towers of fermion loop diagrams.  With a large hierarchy, $M_0/\mu >\!\!> 1$, there
are also large logarithms, and the sum of loop diagrams is best handled
by using an effective action and the renormalization group (RG).
$\Phi(x)$ has only the minimal dynamical degrees of freedom of a pointlike
field.  Hence, the NJL model leads to a pointlike field theory description
of a bound state, with boundary conditions
on the RG running of its couplings at the scale $M_0$.

Since the old top condensation theory used the pointlike NJL
model it therefore omitted an internal bound state 
wave-function, $\phi(r)$.  The formulation of $\phi(r)$, in a
UV completion of the NJL model, has been developed in a recent work \cite{main}\cite{scalars}.
The omission of  $\phi(r)$  
is not expected to significantly affect 
QCD applications of the NJL model, 
due to confinement of quarks which would presumably cut off any wave-function spreading.
However, if the NJL
coupling constant is near its critical value for a non-pointlike
and non-confining theory,
then the
low energy effective theory is approximately conformal. This
implies that the internal
wave-function $\phi(r)$ spreads significantly into empty space.

\vspace{0.0in}
{
\begin{figure}[!htp]
	\centering
	\includegraphics[width=0.5\textwidth]{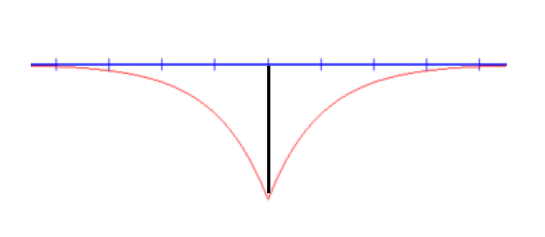}
	\vspace{-0.2in}
	\caption{\normalfont \footnotesize Dirac $\delta$-function potential and its extended wave-function.}
	\label{fig1}
\end{figure}
}

An extended wave function occurs for any 
localized  potential with small eigenvalue for the Hamiltonian. 
The  Dirac $\delta$-function potential in $1+1$ dimensions provides
a typical example.
The Schr\"odinger equation is,
\bbo
-\frac{1}{2m}\frac{d^2}{dx^2}\phi(x)+V(x)\phi(x) =E\phi(x),
\qquad \makebox{where,}\qquad  V(x) = -\alpha \delta(x).  
\ebo 
The bound state solution ($-\infty < x < \infty$) 
is $\phi(x) = -\sqrt{\alpha m}\exp(-\alpha m |x|)  $, 
with eigenvalue $E=-\alpha^2 m/2$.  
The bound state exists for any  $\alpha>0$, with eigenvalue
$E<0$ (analogous to the Coulomb potential in $1+3$ dimensions).  
Hence the critical coupling, the value of $\alpha$ at which $E=0$, is $\alpha =0$, and the external
wave-function then coincides with the lowest energy $1+1$, non-normalizable, scattering state, 
$\phi(x)\rightarrow (constant)$.
Note that the transition from bound to unbound at $\alpha=0$
is discontinuous (non-analytic) since
the normalization,
is finite (compact) for $\alpha>0$ and divergent
(non-normalizable) for $\alpha<0$ \cite{Weinberg}. 
This illustrates the general result that 
a near--critical
bound state in a localized potential 
must always be an extended object, even if the interaction scale
is a very short distance scale.  

There are many models of bound states with
internal structure, such as  \cite{Feynman}\cite{BagModels}\cite{CJT} to name a few.
We prefer, however,  to focus on the well-defined NJL model \cite{NJL} and generalize it to a 
non-pointlike theory.
To understand the internal $\phi(r)$ in field theory
we must first ask exactly how the 
local pointlike 4-fermion interaction is generated as the limit of a  bilocal 
interaction,
$ V(x,y)\rightarrow V(x)$.
 With a bilocal interaction we must then replace the pointlike $H(x)$ by a {\em bilocal field }
$H(x,y)$. In the NJL model the natural candidate for this is ``topcolor'' \cite{Topcolor} where the 
 non--pointlike 
interaction arises from the exchange of a massive gluon-like object,
of mass $M_0$ and coupling $g_0$, called a ``coloron'' \cite{NSD},\cite{Simmons}.
In the large $M_0$ limit
the interaction recovers the pointlike NJL form, but the bilocal nature
of $H(x,y)$ is 
then established and remains so even in the pointlike $ V(x,y)\rightarrow V(x)$ limit!

This requires a general formulation
of a bilocal field theory.  The starting point for this
 begins with old ideas of Yukawa \cite{Yukawa} of multilocal fields.
We can modify and extend Yukawa's bilocal fields to an action formalism.
We then have  a non-pointlike UV description of the physics as a generalization 
of the NJL model which we can rely upon for intuition. The semiclassical binding interaction
is enhanced by a factor of $N_c$ in analogy to BCS superconductivity \cite{BCS}.
The formalism then leads to a
Schr\"odinger-Klein Gordon (SKG) equation that determines $\phi(r)$ with eigenvalue $\mu^2$.
Above critical coupling the eigenvalue,   $\mu^2$, 
becomes negative and spontaneous symmetry breaking (SSB) occurs.

In the symmetric phase of the model (before SSB) we find that
bound states will form semiclassically, $({\cal{\hbar}})^0$, similar to 
the hydrogen atom, but relativistically in a configuration space picture. The
resulting bound state near criticality, where the mass $|\mu|$ is small  compared
to the composite scale (the coloron mass $M_0$), 
will have an extended ``tail'' in its rest frame  $\phi(r)\sim e^{-|\mu|r}/r$
where $r=|(\vec{x}-\vec{y})/2|$.  

The major implication  is that the wave-function spreading 
causes a significant ``dilution'' of $\phi(0)\sim \sqrt{|\mu|/M_0}$. 
The resulting top quark Yukawa coupling, $g_Y\propto \phi(0)$, and quartic coupling,
$\lambda\propto |\phi(0)|^4$, are then determined by
$\phi(0)$, with its power law suppression, rather than the relatively slow RG evolution
in the NJL model. 
By fitting the top quark Yukawa coupling to its known value $g_Y\approx 1$, and $\mu^2=-(88)$ GeV$^2$, the Lagrangian mass 
of the BEH boson in the symmetric phase of the SM,
we can then readily determine $M_0$.
The implied scale of compositeness of $H(x,y)$, i.e., the mass of the coloron, is then 
significantly reduced compared to the NJL model and we obtain 
a result: $M_0\sim 6$  TeV(!).  The 
SM parameters (Lagrangian BEH mass, $-\mu^2$, electroweak VEV, $v_{weak}$,
Yukawa coupling, $g_Y$, and (remarkably) the quartic coupling, $\lambda$) all
become concordant with experiment. Moreover, the 
fine--tuning of the model is vastly reduced to  a few $\%$. The major prediction is the existence of a 
QCD color
octet of colorons with mass $M_0\sim 6$  TeV
that may be accessible to the LHC.\footnote{See Appendix \ref{wallet} 
for a summary of the symmetric and broken phase parameters of the SM. We emphasize that we obtain  results {\em in the semiclassical limit.} We have not yet completed analysis of the quantum loop
corrections to this \cite{CTHprep}, which may be significant, so we  quote $M_0\sim 6$ TeV with
potential uncertainties. }

A core issue of a bilocal (or multilocal) description is the ``relative time'' problem
\cite{Dirac}.
In a two body bound state each particle carries its own clock, hence we have times
$t_1$ and $t_2$, therefore we have the ``average time'' $(t_1+t_2)/2$ and the ``relative time ''
$(t_1-t_2)$.  This is endemic to non-relativistic, as well as relativistic
systems. In the center of mass frame, which is the rest frame of the bound state 
(often called the ``barycentric frame'') the relative time drops out of the kinetic terms.  We can then integrate out the relative time and the interaction 
becomes a static potential in the rest frame. This requires a normalization of
the bilocal field kinetic terms to establish the relevant normalized currents and charges
when relative time is removed. The pointlike NJL theory avoids the relative
time problem because it simplifies the interaction to a single point in spacetime, but one then misses
the extended wave function $\phi(r)$.  For a relativistic
system the reduction is done with Lorentz invariant constraints. While one loses
{\em manifest } Lorentz invariance in the rest frame of the bound state, 
the overall Lorentz invariance of the theory is maintained, a procedure akin to gauge fixing in
a gauge invariant field theory 
(see Appendices \ref{LI} and \ref{CV} for further discussion.)

Mainly, we  propose in the present paper a new version of the top condensation
idea, a ``redux,'' which relies on a ``topcolor'' interaction 
that generates the UV completion of the NJL scheme and provides the binding mechanism
through a bilocal interaction $V(x,y)$.  Though topcolor
was previously introduced in the 1990's, much of its structure carries over 
for us presently \cite{Topcolor},\cite{Topcolor2}.  Here we are invoking it as the  primary binding
mechanism of the BEH boson (replacing, e.g., ``technicolor,'' rather than ``assisting technicolor''). 

We  begin with a quick summary of
key features in the old top condensation NJL model.  We then give a simple example of a bilocal formulation
of the non-relativistic hydrogen atom, which illustrates the formal issues and the problem of relative time. To provide orientation, we then follow with a lightning summary
of the composite BEH theory.  The full technical details, some of which we think are
rather stunning, are
then given in the bulk of the paper.

\subsection{Nambu--Jona-Lasinio Model Application to Top Condensation}

We will rely heavily on intuition from the NJL model, so
we provide this quick summary (more details appear in  \cite{main},\cite{scalars}).
The ``old'' NJL model of top condensation assumes chiral fermions,  with $N_c=3$ ``colors'' and a pointlike 4-fermion interaction, hence we have: 
\bbo
\label{0NJL1}
S_{NJL} 
=\!\int\! d^4x \;\bl i[\bar{\psi}_L(x)\slash{D}_L\psi_{L}(x)]
+ i[\bar{\psi}_R(x)\;\slash{D}_R\psi_{R}(x)]
+\;
\frac{g_0^2}{M_0^2}
[\bar{\psi}_{iL}(x)\psi_{R}(x)]\;[\bar{\psi}_R(x)\psi^i_{L}(x)]\br,
\ebo
where $i$ is an isospin index, 
$[..]$ implies color singlet combination, and $\psi_{R,L}= (1\pm \gamma^5)\psi/2$.

The NJL interaction is invariant under $SU(3)_{QCD}\times SU(2)\times U(1)$ gauge symmetry.
The fields and covariant derivatives are defined in the standard model (for simplicity we don't display the color
indices on the quark fields):
\bbo
\backo\backo
 \psi^i_L= \frac{(1-\gamma^5)}{2}\bl\begin{array}{c} t\\ b \end{array}\br,
 \qquad  \psi_R=\frac{(1+\gamma^5)}{2}t , 
 \nonumbo
 \backo\backo
D_{L\mu}=\partial_\mu- ig_2W^A_\mu \frac{\tau^A}{2}-ig_1 B_\mu \frac{Y_L}{2}
-ig_3{\cal{G}}^A\frac{\chi^A}{2},
 \qquad 
D_{R\mu}=\partial_\mu-ig_1 B_\mu \frac{Y_R}{2}-ig_3{\cal{G}}^A\frac{\chi^A}{2},
\ebo 
where the QCD gluons are ${\cal{G}}^A$, the weak hypercharges
$Y_L= 1/3$, $Y_{tR}=4/3$, and  $Y_{bR}=-2/3$, and the electric charges are as usual: $Q=I_3 + \frac{Y}{2}$
(eg., for the $b_L$ quark, $Q= -\half +\half\frac{1}{3} = -\frac{1}{3}$, while for $b_R$, $Q= 0 -\half\frac{2}{3}
= -\frac{1}{3}$,
etc.).

An equivalent form of the interaction can be written by introducing an auxiliary isodoublet
field $H^i(x)$:
\bbo
\label{0NJL2}
\backo\backo
S_{NJL}
=\!\int\! d^4x \;\bl i[\bar{\psi}_L(x)\slash{D}_L\psi_{L}(x)]+ i[\bar{\psi}_R(x)\slash{D}_R\psi_{R}(x)]
- M_0^2H^\dagger(x)H(x) + (g_0[\bar{\psi}_{iL}(x)\psi_{R}(x)]H^i(x)+h.c.) \br .
\ebo 
The
``equation of motion'' for $H^i(x)$ is then:
\bbo
\label{factor}
 M_0^2H^i(x) = g_0[\bar{\psi}_R(x)\psi^i_{L}(x)].
\ebo
$H(x)$ will become the bound state field.
Note that $H(x)$ is a pointlike field since the 4-fermion interaction is pointlike.

Following  Wilson \cite{Wilson} we view eqs.(\ref{0NJL2}, \ref{factor}) 
as the effective action at the high scale $m=M_0$.
We 
integrate out the fermions to obtain the effective action for the bound state  
field $H(x)$ at a lower scale $m <\!\!<M_0$ :
\bbo
\label{0NJL3}
\backo
S_{\mu}
=\!\int\! d^4x \;\bl i[\bar{\psi}_L\slash{D}_L\psi_{L}]+ i[\bar{\psi}_R\slash{D}_R\psi_{R}]
+
Z D_{H\mu} \Phi^\dagger D_{H}^\mu \Phi
-\mu^2H^\dagger H - \frac{\lambda }{2}(H^\dagger H)^2 + (g_0[\bar{\psi}_{iL}\psi_{R}]H^i(x)+h.c. )
\br.
\ebo
where,
\bbo
\label{0NJL4}
\backo
\mu^2 = M_0^{2}\!-\!\frac{g_0^{2}N_{c}}{8\pi ^{2}}M_0^{2},
\qquad
Z=\frac{g_0^{2}N_{c}}{8\pi ^{2}}\ln(M_0/m), \qquad
\lambda=\frac{g_0^{4}N_{c}}{4\pi ^{2}}\ln( M_0/m).
 \ebo
We see, from Feynman loops, that $H(x)$ acquires a kinetic term with the covariant derivative,
 \bbo
 D_{H\mu}=\partial_\mu- ig_2W^A_\mu \frac{\tau^A}{2}-ig_1 B_\mu \frac{Y_H}{2},
 \ebo
 where the gluons cancel, and the weak hypercharge becomes
 $Y_H= -1$, apropos the BEH boson of the SM.
 
\begin{table}
 \centering
	\includegraphics[width=0.8\textwidth]{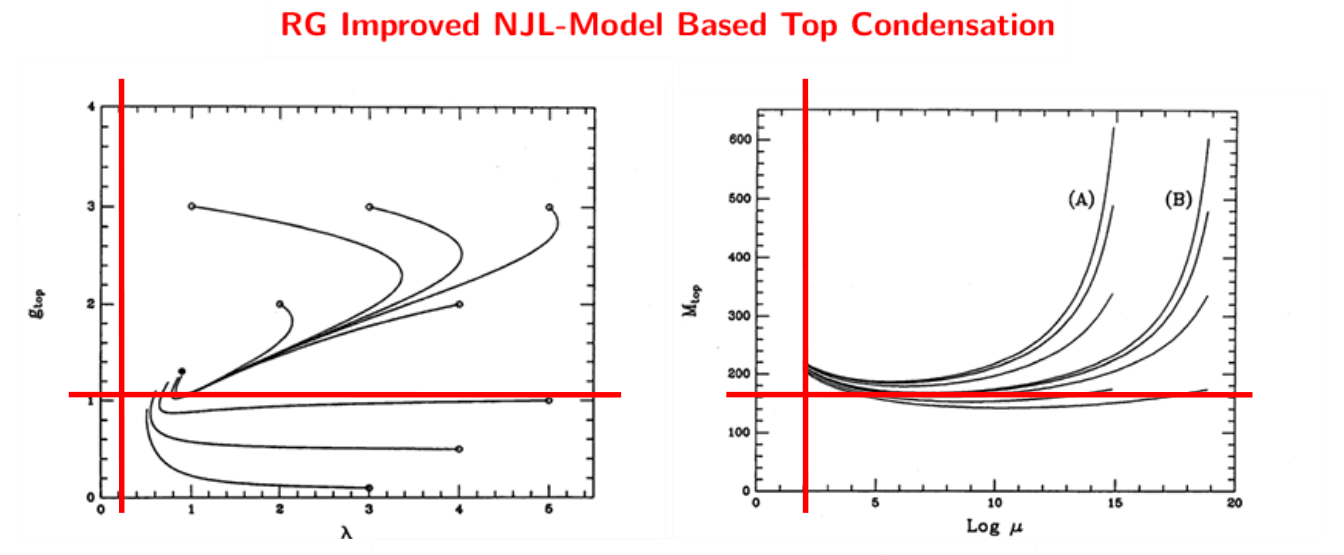}
	\vspace{0.0in}
	
    {{\bf\large Figure 1:} \normalfont \footnotesize  Figure on left 
    shows the full joint RG running of $g_{top}\equiv g_{Y(ukawa)}$ and $\lambda$ flowing  from initial values at
      $M_0= 10^{15}$ GeV to $v_{weak}$.
      Right figure shows full running of effective top quark mass and RG fixed point \cite{PR}.  Solid (red) lines indicate experimental values.}
	\label{fig00}
 \vspace{0.1in}
{\begin{center}
{\renewcommand{\arraystretch}{1.5}
\begin{tabular}{|l|c|c|c|c|c|}
\hline
$M_0$ GeV  & $10^{19}$ & $10^{15}$ & $10^{11} $  & $10^{7}$ & $10^{5} $ \\
\hline
 $m_t$ (GeV) Fermion Loops &  $144$ & $165$ & $200 $  & $277$ & $380$ \\
 \hline
 $m_t$ (GeV) Planar QCD&  $245$ & $262$ & $288 $  & $349$ & $432$ \\
 \hline
  $m_t$ (GeV) Full RG &  $218$ & $229$ & $248 $  & $293$ & $360$ \\
  \hline
   $m_{BEH}$ (GeV)  Full RG &  $239$ & $256$ & $285 $  & $354$ & $455$ \\
   \hline
 \end{tabular}
  \vspace{0.0in}
 \caption{\normalfont \footnotesize  Results for the top quark mass, $m_{top}$, determined
 by RG running from Landau pole in $g_Y$ at $M_0$,
 to $v_{weak}$. }
\vspace{-0.2in}}
\end{center}}	
\end{table}

In particular, note the behavior of the composite BEH boson mass, 
$\mu^2$, of eq.(\ref{0NJL4}) due to
the loop contribution, $ -{g_0^{2}N_{c}M_0^{2}}/{8\pi ^{2}}$, 
(we use a
 UV cut-off  $M_0^{2}$ on the fermion loops to imitate
 a softening of the interaction on scale $m>\!\!> M_0$).
The NJL model therefore has a critical
value of its coupling, $g_c$, defined by the vanishing of $\mu^2$:
\bbo
\frac{g_{c}^{2}N_c }{8\pi ^{2}}=1.
\ebo
We can renormalize, $H\rightarrow \sqrt{Z}^{-1}H$, 
to obtain the full renormalized  effective Lagrangian.  
The notable feature here is that the renormalized couplings
evolve logarithmically in the RG ``running mass'' $m$:
\bbo
\label{0NJL30}
\backo\backo
g^2_Y \!=\! \frac{g_0^2}{Z}\!=\!\frac{4\pi^2}{N_c\ln(M_0/m)},\qquad
\lambda_r\!=\!\frac{\lambda}{Z^2}\! =\!\frac{16\pi ^{2}}{N_c\ln( M_0/m)}.
 \ebo\normalsize
These are the solutions to the RG equations in the large $N_c$ limit, keeping
only fermion loops, 
\cite{CTH}. Eq(\ref{0NJL30}) implies the renormalized couplings have
Landau poles, i.e., $(g^2_Y(m),\lambda_r(m))$ blow up logarithmically
as $m\rightarrow M_0$.  This defines ``compositeness boundary conditions'' on $H$
for the RG running.
We can then use the full RG equations, including QCD and electroweak interactions, 
to obtain precise low energy predictions \cite{CTH}.
In particular, the Yukawa coupling $g_Y$ approaches the IR fixed point value \cite{PR}.
Results are shown in Figure (1) and Table (I).

For  super-critical NJL coupling, $g^2>g^2_c$,
we see that the (renormalized) Lagrangian mass, $\mu_r^2= \mu^2/Z <0$, 
implying there will be a vacuum instability.
The effective action, with the induced quartic coupling  $\sim \lambda_r(\Phi^\dagger\Phi)^2$ term, 
yields the usual sombrero potential, and the $SU(2)\times U(1)$ symmetry is spontaneously broken,
and the neutral component of the  BEH field $H$ acquires a VEV,
 \bbo
\langle H^0\rangle =v_{weak} = \frac{|\mu_r|}{\sqrt{\lambda_r}}.
 \ebo
The top quark then acquires mass $m_t = g_Yv_{weak}$ (for light quark and lepton masses and mixings
we would rely upon higher dimension ``Extended Technicolor'' (ETC) operators \cite{Eichten}). 

The solutions for the NJL based top quark mass are shown in Table I.
At the time the model was proposed there were upper bounds on the
top quark and BEH boson masses of order several hundred  GeV. 
We see that, to obtain a top quark mass $m_t \lta 230$ GeV, 
we require very large $M_0$ due to the slow running of the RG and its fixed point.  
With a choice of e.g.,  of $M_0\sim 10^{15}$ GeV, we obtain
 from eq.(\ref{0NJL4}):
\bbo
\mu_r^2 \sim  M_0^2\bl 1\;-\;\frac{g_0^{2}N_c}{8\pi^{2}}\br = M_0^2\bl 1\;-\;\frac{g_0^{2}}{g^2_{c}}\br.
\ebo
We see that small BEH mass, $\mu^2_r$, mandates the fine--tuning of $g_0^2/g_{c}^2$ at the level of  $ \sim\; 10^{-26} $: 
\bbo
\frac{\delta g_0^{2}}{g_{c}^2}\; \sim  \;\frac{|\mu_r^2|}{M_0^2} \;\sim\; 10^{-26}\; !
\ebo
 
While the top condensation theory was directly testable by experiment, it evidently
failed.  However, the
difficulties with the old top condensation theory
stem from the limitations of the Nambu--Jona-Lasinio model.

\subsection{The Hydrogen Atom as a Bilocal Field Theory \label{hydrogen}}

As a warm-up example we presently give a derivation of the well-known Schr\"odinger
equation for the hydrogen atom, using bilocal field techniques that we develop
for a composite BEH boson.  This illustrates issues that will arise subsequently for the problem of 
bound states of pairs of chiral fermions and our generalization of the NJL model.

We can represent the hydrogen atom in terms
of proton and electron  fields (scalars for present purposes),
$\psi_p(y)$ and
$\psi_e(x)$, where $(x^\mu,y^\mu)$ are 4-vectors. 
We then introduce a bilocal field describing a proton-electron pair,
\bbo
\label{startdef}
\Phi(x,y) = \psi_p(y)\psi_e(x).
\ebo
Since we typically define normalizations as $\int d^3x |\psi|^2 =1$, therefore
the $\psi$ fields have mass-dimension $1/\sqrt{V}\sim M^{3/2}$.  We will presently
assume $\Phi$ has mass dimension $\sim M^{3}$.

A non-relativistic (overall dimensionless) bilocal action can be written as:
\bbo
  \!\!\int \! d^4x \; d^4 y \bl  iZ\Phi^\dagger\frac{\partial}{\partial x^0}\Phi+iZ\Phi^\dagger\frac{\partial}{\partial y^0}\Phi
-\frac{1}{2m_p}Z|\partial_{\vec{y}}\Phi|^2-\frac{1}{2m_e}Z|\partial_{\vec{x}}\Phi|^2 
-e^2D(x-y)|\Phi|^2 \br.
\ebo
We have included a normalization factor, $Z$, of mass dimension $\sim M$ on the bilocal kinetic terms which
will become clear momentarily.
The interaction is generated by a single photon exchange between the non-relativistic
charge densities of the proton and electron, and using eq.(\ref{startdef}):
\bbo
e^2|\psi_e(x)|^2|\psi_p(y)|^2D(x-y) = -e^2 |\Phi(x,y)|^2\int \frac{d^4q}{(2\pi)^4} \frac{e^{iq_\mu (x-y)^\mu}}{q_0^2-\vec{q}^2+i\epsilon}.
\ebo
It is useful to go to coordinates where the heavy proton is located at $X^\mu$ and the
electron at $X^\mu+\rho^\mu$ as,
\bbo
X=y; \;\;\; \rho = x-y;\;\;\;\; \makebox{then,}\;\;\; \partial_x=\partial_\rho;\;\;\;\partial_y = 
\partial_X - \partial_\rho ;
\;\;\;\makebox{with Jacobian:}\;\;\; \left|\frac{\partial(X,\rho)}{\partial(x,y)}\right|\equiv J = 1,
\ebo
and the action becomes, with $\Phi'(X,\rho)=\Phi(X+\rho,X)$,
\bbo \label{hydkt}
\backo\backo 
\int\!\! dX^0d^3X d\rho^0 d^3\rho  
\bl\!  iZ\Phi'{}^\dagger\frac{\partial}{\partial \rho^0} \Phi'+
iZ\Phi'{}^\dagger\bl\frac{\partial}{\partial X^0}-\frac{\partial}{\partial \rho^0} \br\Phi'
-\frac{1}{2m_p}Z|(\partial_{\vec{X}}-\partial_{\vec{\rho}} )\Phi'|^2 
-\frac{1}{2m_e}Z|\partial_{\vec{\rho}}\Phi'|^2 
-e^2D(\rho) |\Phi'|^2\!\br.
\ebo
We see in eq.(\ref{hydkt}) that the derivative $\partial/\partial{\rho^0}$ cancels. 
This is also central to a relativistic formalism: $\rho^0$ is the ``relative time'' and is seen to  drop out of the action
in the rest frame.  The only time degree of freedom then 
 carried by the system is $X^0$ and we can 
 therefore integrate out $\rho^0$.  This reveals the purpose of the normalization
factor $Z$, and we  define:
\bbo
\label{reltimeZ}
\int d\rho^0 \; Z=1.
\ebo
$Z$ acts only on the kinetic terms and is needed to insure these are canonical, e.g., that they generate properly normalized Noether currents, etc.  Note that we could alternatively define
$Z=\delta(\rho^0)$.

We therefore assume that $\Phi$ has no dependence upon $\rho^0$ (this can be done in the relativistic case with Lorentz invariant constraints).  We can then integrate over $\rho^0$
in the interaction which then becomes the standard Coulomb form,
\bbo
\backo e^2\!\!\! \int\!\! d\rho^0  D_F(\rho)|\Phi'(\vec{\rho})|^2= -e^2 \!\!\!\int\!\!  d\rho^0 \frac{d^4q}{(2\pi)^4}\frac{1}{q^2}e^{i q_\mu \rho^\mu }|\Phi'(\vec{\rho})|^2
= e^2\!\!\! \int\!\!   \frac{d^3q}{(2\pi)^3}
\frac{1}{\vec{q}{\;}^2}e^{-i \vec{q}\cdot\vec{\rho}  }|\Phi'(\vec{\rho})|^2=  \frac{ e^2}{4\pi |\vec{\rho}|}|\Phi'(\vec{\rho})|^2.
\ebo 
We can then approximate the proton as infinitely heavy,
since the proton kinetic term is suppressed as $1/m_p\sim 0$.
In the rest frame the proton is a zero-momentum plane wave with ``volume normalization,''  $\psi_p(x)=e^{iEX^0}/\sqrt{V}$
and we can integrate it
out with $\int d^3X=V$.
We then have  $\Phi'(X,\rho) \rightarrow e^{-iEX^0}\psi_e(\vec{\rho}) $, where the ``clock time'' is $X^0=t$, and we have a static electron wave-function $\psi_e(\vec{\rho})$.
The bilocal action then becomes the usual single particle form for the electron:
\bbo
\int \!\! dt\; d^3 \rho\; \bl E|\psi_e(\vec{\rho}\;)|^2 -\frac{1}{2m_e}|\partial_{\vec{\rho}\;}\psi_e(\vec{\rho}\;)|^2
+ \frac{\alpha}{|\vec{\rho}\;|}|\psi_e(\vec{\rho}\;)|^2 \br.
\ebo
In spherical coordinates, integrating by parts and 
extremalizing the action gives the Schr\"odinger equation,
\bbo
-\frac{1}{2m_e}\nabla_{\vec{\rho}\;}^2 \psi_e(\vec{\rho}\;)
- \frac{\alpha}{|\vec{\rho}\;|}\psi_e(\vec{\rho}\;) = E \psi_e(\vec{\rho}\;),
\ebo
where $E$ is the eigenvalue, and $\alpha=e^2/4\pi$. 

This illustrates that the natural starting point for a two-body bound state is
a bilocal field, $\Phi(x,y)$,  
and schematically anticipates the treatment we will use below for pairs of chiral fermions \cite{main}.
To give another less trivial example, we give a relativisitic bilocal field
theory composed of scalar fields in  ref.(\cite{scalars}).
We now turn to the UV completion and bilocalization of the NJL model.

 \subsection{Outline of a Bilocal BEH Boson  Theory}
 
We presently give a brief summary of
our theory of a composite BEH boson 
arising in ``natural'' top quark condensation. 
This will illustrate
the principles of the construction and one will see similarities
with the simple hydrogen atom example given above.
This preliminary summary omits  the technical details that follow in the bulk of the paper.

In the  NJL top condensation model the pointlike
4-fermion effective interaction of eq.(\ref{oneone})  can be viewed as
a Fierz rearrangement of a color-current interaction
(the Fierz rearrangement is derived in Appendix C of \cite{main}):
\beq
  -\frac{g^2}{M_0^2}[\bar{\psi}_{iL}\psi_{R}][\bar{\psi}_{R}\psi_L^i]
=
\frac{g^2}{M_0^2}(\bar{\psi}_{iL} \gamma_\mu \frac{\chi^A}{2} \psi^i_L )
(\bar{\psi}_{R} \gamma^\mu \frac{\chi^A}{2} \psi_{R}) + O(1/N_c),
\eeq
where  $N_c=3$ is the number of colors.  This is exactly
the form (including the sign) induced
by a massive color octet vector boson exchange, and leads
to topcolor models  \cite{Topcolor}.

In topcolor, QCD is embedded into an $SU(3)_1\times SU(3)_2$
gauge group at higher energies.  The second (weaker)
$SU(3)_2$ gauge interactions acts upon the first and second
generation quarks while the (stronger) $SU(3)_1$
interaction acts upon the third generation and
drives the formation of the BEH bound state, $H$. Additional dynamics is also incorporated
to disallow the formation of a second BEH boson, $H'$, containing $b_R$,
(usually achieved by introducing a heavy $Z'$
so the $\bar{\psi}_Lb_R$ channel is subcritical; this can be accommodated in extension of the
present minimal model). In 
the following minimal model we simply omit the $b_R$ quark from the binding dynamics.

The gauge structure 
at high energies (ignoring any $Z'$ interactions)
is therefore \cite{Topcolor}:
\beq
SU(3)_1\times SU(3)_2
\times SU(2)_L \times U(1)_{Y} \rightarrow
SU(3)_{QCD}\times SU(2)_L\times U(1)_{Y},
\eeq
The SM 
fermions are  assigned to $(SU(3)_1, SU(3)_2, SU(2), Y)$, where $Q= I_3+\frac{Y}{2}$, as
follows:
\bbo
(t,b)_L \sim   (3,1,2,{1}/{3}) \qquad \;\;
(t)_R \sim    (3,1,1,{4}/{3})  \qquad
[(b)_R \sim    (3,1,1,-{2}/{3} )] \nonumbo
(\nu_\tau,\tau)_L \sim  (1,1,2,-1) \qquad
(\tau)_R \sim   (1,1,1,-2) \nonumbo
(u,d)_L,\;\;\;  (c,s)_L \;\;\sim   (1,3,2,{1}/{3}) \qquad
(u)_R, \;\; (c)_R \sim \left(1,3,0,{4}/{3}\right) \nonumbo
(\nu_e, e)_L,\;\; (\nu_\mu, \mu)_L  \sim  (1,1,2,-1) \qquad
(\ell)_R, \;\;  (\mu)_R \sim \left(1,1,1,-2\right).
\ebo
The $ SU(3)_1\times SU(3)_2$ extended color interaction is broken to the diagonal $SU(3)_{QCD}$
(this is described elsewhere \cite{Topcolor}\cite{NSD})
leading to the
massive octet of ``colorons,'' ${\cal{G}}^{\prime A}_{\mu}$, and 
the massless octet of the gluons of QCD, ${\cal{G}}_\mu^A$.

Integrating out the heavy colorons and Fierz rearranging
gives  a bilocal interaction:
\bbo
\backo
S'=g_0^2\!\!\int\!\!d^4x d^4y\; [\bar{\psi}_{iL}(x)\psi_{R}(y)] D_F(x-y)[\bar{\psi}_R(y)\psi^i_{L}(x)].
\;\;\;
\ebo 
We therefore introduce a color singlet bilocal BEH field of
mass dimension $1$, (analogous to  eqs.(\ref{factor}, \ref{startdef}):
\bbo
\label{Phi0111}
\sqrt{N_c} M_0^2H^i(x,y)= [\overline{\psi}_{R}(x)\psi^i_{L}(y)], 
\ebo
(here $(i,j)$ are electroweak indices, $[..]$ denotes color indices summed,
and $H^i$ is conventionally color normalized, as in Section \ref{bilocalfield} below).
The fields appearing on the {\em rhs} of eq.(\ref{Phi0111}) are those that will form 
the bound state when the interaction is turned on, generally the low momentum scattering states.
The interaction then becomes bilocal,
\bbo \backo
 g_0^2M_0^4 N_c \!\!\int\!\! d^4x d^4y \; D_F(x-y) |H(x,y)|^2,
\;\;\;\makebox{where,}\;\;\;
D_F(x-y) =- \int\frac{1}{q^2-M_0^2}e^{iq(x-y)}\frac{d^4q}{(2\pi)^4},
\ebo 
where, due to a color singlet normalization of $H$,
an $N_c$ enhancement occurs in analogy to BCS theory \cite{BCS} 
(as detailed in Section \ref{semiclass}).
The interaction also generates Yukawa couplings
of $H$ to fields that remain free fermions,
\bbo
 g_0^2M_0^2\sqrt{N_c}\!\!\int\!\! d^4x d^4y \;
D_F(x-y) [\bar{\psi}{}^i_{L}(y)\psi_{R}(x)]_f H_i(x,y) +h.c.,
\ebo 
We can then construct the Lorentz invariant action that yields the equations of motion by variation:
\bbo 
\label{first}
\backo
S_K=
M_0^4 \int d^4x d^4y \; \bl Z|D_R^\dagger H(x,y) |^2
+ Z|D_L H(x,y)|^2 +{g_0^2 N_c}D_F(x-y)|H(x,y)|^2 \br,
\ebo
where the covariant derivatives are as defined in the NJL model:
\bbo
 \backo\backo
D_{L\mu}=\frac{\partial}{\partial y_\mu}- ig_2W(y)^A_\mu \frac{\tau^A}{2}-ig_1 B(y)_\mu \frac{Y_L}{2};
 \qquad 
 D^\dagger_{R\mu}=\frac{\partial}{\partial x_\mu}+ig_1 B(x)_\mu \frac{Y_R}{2},
\ebo 
Note that $D_L$ ($D^\dagger_R$) acts at coordinate  $y$ ($x$), and $D^\dagger_R$ acts on $\bar{t}_R$,
hence the sign flip in the gauge field terms (note the derivative $D^\dagger$ acts in the forward
direction as we have written the kinetic term eq.\ref{first}).

We now pass to barycentric coordinates,
\bbo
\label{bary}
\backo\backo
X^\mu=\frac{x^\mu+y^\mu}{2},\;\;\; r^\mu=\frac{x^\mu-y^\mu}{2}, 
\;\;\;
\partial_x=\half(\partial_X+\partial_r),
\;\;\;
\partial_y=\half(\partial_X-\partial_r)
\;\;\;\makebox{with Jacobian:}\;\;\; \left|\frac{\partial(X,\rho)}{\partial(x,y)}\right|\equiv J = 2^4.
\ebo
We then use Wilson lines to ``pull-back''
the gauge couplings from $(x,y)$ to the center $X$. This is
done by field redefinitions (as discussed in Appendix \ref{gauge}):
\bbo
H(x,y)\rightarrow W^\dagger_R(X,x)W_L(X,y) H(X)\phi(r),
\ebo
where we have also made a factorized ansatz for the $H^i$ field following Yukawa \cite{Yukawa}:
\bbo
H^i(x,y) \rightarrow \sqrt{2/J}\; H^i(X)\phi(r),  \qquad \makebox{where $\phi$ is normalized as,}
\qquad ZM_0^4 \int d^4 r |\phi(r)|^2 = 1,
\ebo
and $\phi$ is therefore dimensionless.
This leads to the action:
\bbo
\label{Haction}
=
 \!\!\int\!\!d^4X\; \bl |D_{H}H(X) |^2
+|H(X)|^2 M_0^4  \int \!\!d^4r\bl 
Z |\partial_{r^\mu}\phi(r) |^2 +2{g_0^2 N_c}D_F(2r)|\phi(r)) |^2\br \br,
 \ebo      
where the $H^i(X)$ kinetic term is now canonical, and the covariant derivative is:
 \bbo
 D_{H\mu}=\frac{\partial}{\partial X^\mu}- ig_2W^A(X)_\mu \frac{\tau^A}{2}-ig_1 B(X)_\mu \frac{Y_H}{2}.
 \ebo
 We see that the ``pull-back'' of the Wilson lines has moved all the electroweak gauging
 of the bilocal field to the ``center,'' $X$, and changes the covariant derivative 
to the precise form apropos the BEH boson, (and the gluons have cancelled).  
The BEH boson has therefore become
a ``hedge-hog'' configuration of radiating Wilson lines. The internal wave-function $\phi(r)$
is now a complex scalar that carries no gauge charges. 

We note that the 
action is Lorentz invariant (see Appendix \ref{LI})  and can therefore
be evaluated in any frame. If we consider a pair of massless particles of 4-momenta, $p^\mu_1$ and $p^\mu_2$,
we have total momentum, $P^\mu=(p^\mu_1+p^\mu_2)$, and relative momentum,  $Q^\mu=(p^\mu_1-p^\mu_2)$, where
$P_\mu Q^\mu=p_1^2-p_2^2=0$. 
This implies that there is always a rest frame in which
$P^\mu=(P_0,0)$ and $Q^\mu=(0,-\vvq)$. 
In the rest frame we see that $Q^0=0$, therefore the relative time, $r^0$,
drops out of the kinetic terms.

Hence, we can assume $\phi(r)\rightarrow \phi(\vec{r})$ 
is  a ``static field,'' with no dependence on $r^0$.
This converts the
$\partial_{r^\mu}^2\rightarrow -\partial_{\vec{r}}^2$. 
We then define $Z$ by 
$ZM_0\int dr^0 =1$  in analogy to
eq.(\ref{reltimeZ}),  (or $Z\rightarrow \delta(M_0\omega^\mu r_\mu)$, 
where $\omega_\mu = P_\mu/\sqrt{P^2}$ is a timelike unit 4-vector).
The $\phi$ normalization becomes:
\bbo
1= \int \!\! d^3r\; M_0^3\;|\phi(\vvr)|^2.
\ebo
The coloron exchange potential when integrated over $r^0$ then
becomes a static Yukawa potential.
Including the Yukawa interaction 
and a loop generated quartic term (see Section \ref{all}), we obtain
the effective action for the composite BEH field,
  \bbo
\backo\backo S=
\!\!\int\!\! d^4X\;\bl
 |D_H H(X)|^2+ 
|H(X)|^2 \; M_0^3\!\! \int\!\! d^3r\; \bl
-|\partial_{\vec{r}} \phi(r)|^2 
+{g_0^2 N_c M_0}\frac{ e^{-2M_0 |\vvr|}}{8\pi |\vvr|}|\phi(\vvr) |^2\br 
\nonumbo\qquad\qquad
-\frac{\lambda}{2}(H^\dagger H)^2 - g_Y\left([\bar{\psi}_{iL}(X)t_{R}(X)]_{f}H^i(X)  +h.c.\right)\br.
   \ebo   
The internal field $\phi(r)$ is ``nested'' within the
action for a conventional pointlike BEH boson, $H(X)$. The static $\phi$ field
has Hamiltonian:
\bbo
{\cal{M}}=
M_0^3\!\! \int\!\! d^3r\; \bl
|\partial_{\vec{r}} \phi(r)|^2 
-{g_0^2 N_c M_0}\frac{ e^{-2M_0 |\vvr|}}{8\pi |\vvr|}|\phi(\vvr) |^2\br .
\ebo
Extremalization of this yields the Schr\"odinger-Klein-Gordon (SKG) equation
for $\phi$ with the eigenvalue $\mu^2$:
 \bbo
 -\nabla^2 \phi -g_0^2N_c M_0\frac{ e^{-2M_0 |\vvr|}}{8\pi |\vvr|}\phi(r) =\mu^2\phi.
 \ebo
We find that the SKG equation has
a critical coupling, $g_c$, for which $\mu^2=0$,  that is very close to the quantum
NJL critical coupling (see Sections \ref{exact2} and \ref{numerical}).  When $g_0>g_c$ the eigenvalue  $\mu^2$
becomes negative, $=-|\mu|^2$.  In such a solution the action
for $H(X)$ then becomes the familiar,
\bbo
\backo\!\!\! S=
\!\!\int d^4X \bl
 |D_H H(X)|^2+|\mu|^2 |H(X)|^2
-\frac{\lambda}{2}(H^\dagger H)^2 - g_Y\left([\bar{\psi}_L(X)t_{R}(X)]_{f}H(X)  +h.c.\right)\br,
   \ebo 
 with the ``sombrero potential'':
 \bbo
-|\mu|^2 |H(X)|^2 +\frac{\lambda}{2}(H^\dagger H)^2 .
   \ebo 
 The technical details of this derivation are many,  and together with the analysis of the results, are given
 in the remainder of this paper. The results of the present scheme are:
 
 \begin{itemize}
  \item
  The solution of the SKG equation for $\phi(r)$ indeed extends to large distances, $\phi(r) \sim e^{-|\mu|r}/r$ where $|\mu|<\!\!M_0$ near critical coupling.
  This dilutes the value of $\phi(0) \sim \sqrt{|\mu|/M_0}$. 
  We find that the Yukawa coupling $g_Y \propto \phi(0)$
  and $\lambda \propto g_Y^4 \propto  |\phi(0)|^4$.
  This has profound effects on the theory compared to the pointlike NJL model.
  
  \item Inputting the known value of the Lagrangian mass of the BEH boson
  in the symmetric phase, which is 
  $-|\mu|^2 =-(88)^2$ GeV$^2$, we find that {\bf the scale $M_0$ is now
  $M_0\approx 6$ TeV } (cf, no longer the nonsensical $10^{15}$ GeV in the NJL model).
  
  \item Moreover, the quartic coupling, $\lambda$,  is now determined at loop level by RG
  running from $M_0$, with boundary condition $\lambda=0$ (not a Landau pole!) 
  down to $|\mu|\sim 88$ GeV.
  This yields, at one loop,  $\lambda \approx 0.23$, whereas the standard model
  determines $\lambda \approx 0.25$, hence remarkable agreement is obtained
  (whereas in the NJl model we had $\lambda\sim 1$).
  
  \item The degree of fine-tuning of the theory is also suppressed by $\phi(0)$ in
  a subtle way. Rather than the naive result one would expect from the NJL model, $\delta g_0^2/g_c^2 \sim|\mu|^2/M_0^2 \sim 10^{-4}$,
  we now obtain a linear relation: $\delta g_0^2/g_c^2 \sim|\mu|/M_0 \sim 1\%$.
   
 \end{itemize}

  This concludes a lightning summary to give the reader a sense of 
  this approach and what it yields.  We now descend into the technical details.

\newpage
 
\section{Semiclassical Non-Pointlike Generalization of the NJL Model \label{semiclass} }

\subsection{Bilocal Fields \label{bilocalfield}}

We now consider in greater detail the  formalism for a semiclassical approach to binding in a non-confining theory
of chiral fermions in analogy to our brief sketch of the hydrogen atom above.

In the limit of shutting off an interaction, a bound state
is just a two-body scattering state, such as a product of a free electron 
and free proton wave-functions in the case of hydrogen.  For chiral fermions this can
be described by a complex bilocal field $\Phi^A_B(x,y)$, 
\bbo
\label{Phi01}
M^2\Phi^A_B(x,y)=\overline{\psi}^A_{R}(x)\psi_{BL}(y),
\ebo
(where $(A,B)$ arbitrary unsummed color and flavor indices for more general $G_L\times G_R$ chiral group).
Note that here we have implicitly defined $\Phi$ as a mass dimension-1 field, like a scalar,
 and the mass prefactor, $M^2$, will be elaborated below.
$\Phi$ 
represents a ``bosonization'' of the pair of chiral fermions, as is done in 
writing  chiral Lagrangians, such as the $\Sigma$-model.
$M$ is {\em a priori} arbitrary, but will be determined dynamically.
$\Phi$ can in principle describe arbitrary pairs
of fermions, including bound states or open scattering states.

 Eq.(\ref{Phi01}) has a formal similarity to the factorized auxiliary field of the NJL model
 in eq.(\ref{0NJL2}), however, $\Phi^{A}_{B}(x,y)$ is now a distinct physical free field. 
 Unlike the auxiliary field in the NJL model
 it's kinetic term is not induced by loops, and it will have a free field kinetic term.
 We'll presently restrict ourselves to a single flavor, hence a $U(1)_L\times U(1)_R$
 flavor symmetry, and $(A,B)\rightarrow (a,b)$
 are $SU(N_c)$ color indices (this can be readily extended to $G_L\times G_R$ flavor group).

In the UV completion (coloron) model of the next section, we will see that only the color singlet field
forms a bound state of a pair of chiral fermions.
With $SU(N_c)$ color indices, $(a,b)$, the field
$\Phi^{a}_{b}(X,r)$ is a complex matrix that transforms as a product
of $SU(N_c)$ representations,  $\bar{N}_c\times  {N}_c$, and
therefore decomposes into a singlet plus an adjoint representation.
We designate the color singlet bilocal field
as $\Phi^0$ and conventionally normalize it as,
\bbo
\label{singlet}
{\Phi}^{a}_{b}(x,y) =  \frac{1}{\sqrt{N_c}}\delta^a_b\Phi^0(x,y).
\ebo\normalsize
The conventional normalization allows canonically normalized kinetic terms,
$\Tr[\partial \Phi^\dagger \partial \Phi]
= \partial \Phi^0{}^\dagger\partial\Phi^0$.
Note $\Tr \Phi = {\Phi}^{a}_{a}(x,y) =  {\sqrt{N_c}}\Phi^0(x,y)$.

We can have both bound and unbound
free fermionic two-body scattering states (these are modes that will remain free after the
interaction is turned on), and we will denote the  free fermion pair
by $\overline{\psi}^A_{R}(x)\psi_{BL}(y)_f$
with subscript $f$. 
We can therefore consider a quantum state consisting
of superposition of a to-be-bound state and to-remain-free fermions written as,
\bbo
 \label{Phi02}
 \bar\psi{}^A_R(x) {\psi}_{LB}(y)= \bar\psi{}^A_R(x) {\psi}_{LB}(y)_{f} + M^2\Phi^{A}_{B}(x,y).
 \ebo\normalsize 
Technically, the components of this are orthogonal,
 \bbo
 \label{0Phi01}
 \int\!\!d^4x d^4y\;\bar\psi{}^A_R(x) {\psi}_{LB}(y)_{f}\Phi^\dagger{}^{A'}_{B'}(x,y)=0,
 \ebo
 which would become relevant when we do perturbation theory.
Since only the color singlet binds, we can rewrite eq.(\ref{Phi02})
containing free fields and the color singlet bound state of eq.(\ref{singlet}),
 \bbo
\label{act2}
\backo\!\!\!\!\!\!\! \bar{\psi}^{a}_L(x)\psi_{bR}(y)
\rightarrow  \bar{\psi}^{a}_L(x)\psi_{bR}(y)_{f}
+ M^2\frac{\delta{^a_b}}{\sqrt{N_c}} \Phi^{0}(x,y).
 \ebo

\subsection{The Coloron Model $\label{3}$}

As described above the pointlike NJL model can be viewed as the limit of a physical theory
with a bilocal interaction.
The primary example  is the   ``coloron model'' 
\cite{Topcolor,NSD,Simmons}.
The coloron is a perturbative, massive gauge boson, a massive  analogue
of the gluon, arising in a 
local $SU(N_c)$ gauge theory broken to a global $SU(N_c)$.

We integrate out the massive coloron to generate a single particle exchange
potential that defines the model. This leads to a bilocal current-current form:
\bbo
\label{TC0}
\backo
\;S'\!=\!-
{g_0^2}\!\!\int\!\! d^4x d^4y\; 
 [\bar{\psi}_{L}\!(x)\!\gamma_\mu T^A \psi_{L}\!(x)]D^{\mu\nu}(x-y)
[\bar{\psi}{}_{R}(y)\! \gamma_\nu T^A \psi_{R}(y)],
\ebo
where $T^A=T_a^{Ab}$ are generators of $SU(N_c)$, and color indices are contracted within
brackets $[...]$.

The coloron propagator in Feynman gauge is:
\bbo
\label{propagator}
D_{\mu\nu}(x-y)= g_{\mu\nu} D_F(x-y);
\qquad
D_F(x-y) =- \int\frac{1}{q^2-M_0^2}e^{iq(x-y)}\frac{d^4q}{(2\pi)^4}.
\ebo\normalsize
A Fierz rearrangement of the interaction to leading order in $1/N_c$ 
leads to a potential:
\bbo
\label{coloronexchange}
\backo
S'=g_0^2\!\!\int\!\!d^4x d^4y\;[\bar{\psi}_L(x)\psi_{R}(y)] D_F(x-y)[\bar{\psi}_R(y)\psi_{L}(x)],
\ebo\normalsize 
(the Fierz rearrangement is given explicitly in Appendix C of ref.\cite{main}).

$S'$  of eq.(\ref{coloronexchange}) is the most attractive channel and leading
in large $N_c$.
Hence, we
replace the pointlike 4-fermion interaction with 
the non-pointlike  $S'$ of eq.(\ref{coloronexchange}).
Note that if we suppress the $q^2$ term in the denominator,
of eq.(\ref{propagator}) we have,
 \bbo
\label{4NJL}
{D}_F(x-y)\rightarrow  \frac{1}{M_0^2}\delta^4(x-y),
 \ebo
and  we recover the pointlike NJL model interaction, corresponding
to the large $M_0^2$ limit.

Now,
substitute eq.(\ref{act2})  into eq.(\ref{coloronexchange}) to obtain,
\bbo 
\label{5NJL}
\backo\backo
S'\;
\longrightarrow\;
g_0^2 \!\int\!d^4x\; d^4y\;[\bar{\psi}_L(x)\psi_{R}(y)]_f D_F(x-y) [\bar{\psi}_R(y)\psi_{L}(x)]_{f}
\nonumbo
+\;{ g^2_0\sqrt{N_c}}M^2\!\!\int\!\!d^4x\; d^4y\;[\bar{\psi}_L(x)\psi_{R}(y)]_{f}  D_F(x-y)\;\Phi^0(x,y) {+h.c.}
\nonumbo
+\; { g_0^2N_c} M^4\!\!\int\!\!d^4x\; d^4y\; \Phi^0{}^\dagger(x,y)\; D_F(x-y)\;\Phi^0(x,y).
\ebo
The leading (first) term $S'$ of eq.(\ref{5NJL}) is the unbound  4-fermion
scattering interaction 
and has the structure of the NJL interaction in the limit of eq.(\ref{4NJL}) and
identifies $g_0$ as the analogue of the NJL coupling constant.
The second term, $\sim g^2_0\sqrt{N_c}[\psi^\dagger\psi]D_F\Phi^0+h.c.$,  
has the form of the Yukawa interaction between
the bound state $\Phi^0$ and the free fermion scattering states.
Note the appearance of the 
color factors, $\sqrt{N_c}$ and $N_c$, in the second and third lines respectively. 
The third term $\sim D_F|\Tr\Phi|^2 \propto N_c$, is the potential that makes the
semiclassical bound state, with the $N_c$ enhancement
analogous to a BCS superconductor \cite{BCS}.

In this scheme, the mass scale, $M$, is ultimately inherited by the bound state $\Phi$ from the coloron interaction
which introduces the scale $M_0$. 
The prefactor, $M$, in eq.(\ref{Phi01}), should be viewed part of the wave-function
of $\Phi$. It
is {\em a priori} arbitrary, but
we can swap it for a dimensionless 
 parameter $\epsilon$ as:
\bbo
 M=\epsilon M_0.
  \ebo
 The theory, like the NJL model, is viewed as having
  a maximum cut-off mass scale, $M_0$, hence $\epsilon \leq 1$ (since, for 
  $q^2 >\!\!> M_0^2$, the interaction turns off as $1/q^2$). 
 In a variational
 calculation of the effective potential (in Section \ref{variational} below) we will see that $\epsilon$ is determined as a
 minimum.  We find that $\epsilon =1$ 
 extremalizes the SKG effective potential in a bound state with super critical
 coupling $g_0^2 > g_c^2$, and generates a negative eigenvalue $\mu^2$.
 Furthermore, we find that
 $\epsilon =0$ is the extremal value for the subcritical case. 
 Hence in the subcritical case, $\Phi$, as a stable bound state, disappears and we are left
 with only unbound fermions, while the bound state with negative $\mu^2$ will
 lead to spontaneous symmetry breaking.  
 Note $\epsilon$ rescales the coupling constant
 $\sim g_0^2\epsilon$ and the maximal strength of the interaction for a given
  $g_0^2$
corresponds to $\epsilon \rightarrow 1$.

\subsection{Fake Chiral Instability: The Need for the
Dynamical Internal Wave Function \label{instab}}

Consider the pointlike limit of eq.(\ref{5NJL}) 
and the semiclassical fields in eq.(\ref{4NJL}), 
where we take a pointlike limit of the potential and  of the wave-function, to replace $\Phi^0(x,y)\rightarrow \Phi^0(x)$,
and obtain,\footnote{ $\Phi(x)$ is analogous to, but {\em should not be confused with the factorized NJL interaction.}, eq.(\ref{0NJL2}), where $M_0^2$ is a right-sign non-tachyonic mass. 
Here $\Phi$ is not a pure auxiliary field, but rather
is physical. }
\bbo
\label{60NJL}
\backo
S'
\rightarrow \!\!\int\!\!d^4x \;\bl
\frac{g_0^2}{M_0^2}[\bar{\psi}_L\psi_{R}]_f [\bar{\psi}_R\psi_{L}]_{f}
+ \widehat{M}^2 \Phi^0{}^\dagger\Phi^0
+ g^2_0\epsilon \sqrt{N_c}([\bar{\psi}_L\psi_{R}]_{f}\Phi^0 {+h.c.})
 \br,
\ebo
where  $\widehat{M}^2=g_0^2N_cM^2$.
Eq.(\ref{60NJL})  
contains a wrong-sign (``tachyonic'') mass term,
implying a potential  $\sim  -\widehat{M}^2|\Phi|^2$.
This appears to generate spontaneous symmetry breaking
for any values of the underlying parameters $M_0$ and $g_0$
and the vacuum implodes.
A chiral vacuum instability is 
apparently an immediate, large effect of introducing eq.(\ref{act2})!

Such a conclusion is obviously
physically incorrect.  In
naively replacing $\Phi(x,y)$
with $\Phi(x)$ we have 
neglected the kinetic term of the internal wave-function, $|\partial_r\Phi|^2$
where  $r=(x-y)/2$.  This
opposes the instability like a repulsive interaction and will
stabilize the vacuum in weak coupling.  This is similar to the
stabilization of the classical hydrogen atom by the Schr\"odinger wave-function.
A chiral instability can occur through competition of the repulsive internal wave-function
kinetic term and the attractive potential,
but will require a sufficiently large coupling, $g_0^2>g_{c}^2$,
to drive it, and a quartic coupling to stabilize the vacuum.

We therefore must consider the internal dynamics of the non-pointlike bound state.
We note that this in the spirit of ref.\cite{hasenfratz},
as these authors were essentially arguing for a bare kinetic
term of the factorized NJL model of eq.(\ref{0NJL2}).
We are presently arguing for the necessity of all
bare kinetic terms of the bilocal field $\Phi(x,y)$.

\section{Relativistic Bilocal Fields}

We begin by again noting the all--important kinematics of a two particle massless fermion state.
For a pair of massless particles of 4-momenta $p_1$ and $p_2$, we have $p_1^2=p_2^2=0$,
and we can have two-body plane waves,
$\Phi(x,y) \sim \exp(ip_1x+ip_2y)$. We  pass to the total
momentum $P=(p_1+p_2)$ and relative momentum  $Q=(p_1-p_2)$,
and the plane waves become $\exp(iPX+iQr)$ where we define ``barycentric coordinates,'' 
\bbo
\label{bary}
X^\mu=\frac{x^\mu+y^\mu}{2},\qquad\;\; r^\mu=\frac{x^\mu-y^\mu}{2}, 
\qquad 
\partial_x=\half(\partial_X+\partial_r),
\qquad\!\!\!\!\!
\partial_y=\half(\partial_X-\partial_r).
\ebo
Note that $P_\mu Q^\mu=p_1^2-p_2^2=0$. 
This implies that there is always a rest frame in which
$P=(P_0,0)$ and $Q=(0,\vvq)$. 
Hence, in the rest frame the dependence upon $\vec{X}$ and, in particular the relative time, $r^0$, drop out.
If the particles are constituents of a bound state then this is the rest
frame of the composite particle.

To proceed we require the generalized kinetic term of $\Phi(x,y)$
viewed as a bilocal field with an internal wave-function coordinate, $r^\mu$.
A free particle scattering state, $\Phi(x,y)$, 
composed of massless particles, will satisfy \cite{Yukawa}:
\bbo
\partial^2_x\Phi(x,y) + \partial^2_y\Phi(x,y)=0
\qquad \makebox{
or equivalently,}\qquad
\half \partial^2_X\Phi'(X,r) + \half\partial_r^2\Phi'(X,r) =0,
\ebo
where $\Phi'(X,r)=\Phi(X-r,X+r)$.\footnote{Note: A more general discussion, including Lorentz invariant constraints, can be found in Section 2.5 of ref.(\cite{main})}

 The bilocal field $\Phi(x,y)=\overline{\psi}_{R}(x)\psi_{L}(y)$
represents a ``bosonization'' of the pair of chiral fermions, as in chiral Lagrangians.
The equations of motion follow from the square of the free particle
Dirac equations, $(\slash{\partial})_{x}^2\psi_{R}(x)=0$ and $(\slash{\partial})_{y}^2\psi_{L}(y)=0$.
Note that we have chosen to describe the separation as $r$, which is the radius, where 
$ 2r=\rho\equiv (x-y) $ denotes
the separation of the particles. The choice of $r$ leads to more symmetrical
expressions in $r$ and $X$, and (somewhat) suppresses inconvenient factors of $2$.

We can construct an action that yields the equations of motion by variation:
\bbo 
\label{xxNJL}
\backo
S_K=
M^4 \! \int\!\! d^4x d^4y \; \bl Z|\partial_x\Phi |^2
+ Z|\partial_y\Phi |^2 +{g_0^2 N_c}D_F(x-y)|\Phi(x,y) |^2 \br
\nonumbo
=\half J
 M^4 \!\!\int\!\! d^4X d^4r\; \bl Z|\partial_{X}\Phi' |^2
+ Z|\partial_{r}\Phi' |^2 +2{g_0^2 N_c}D_F(2r)|\Phi'(2r) |^2 \br,
 \ebo      
where  $J=| \partial(x,y)/\partial(X,r)|=2^4$ is the Jacobian in passing from $(x^\mu,y^\nu)$
to the barycentric coordinates $(X^\mu,r^\nu)$; the factor $\half\times $ comes from the derivatives,
$(\partial^2_x+\partial^2_y)\rightarrow \half (\partial^2_X+\partial^2_r).$
The normalization factor $Z$ is necessary to remove relative time and maintain canonical 
kinetic terms (as in the hydrogen atom example of Section \ref{hydrogen} above).  
 
Following Yukawa \cite{Yukawa}, consider the factorized ansatz of  $\Phi'(X,r)$:
 \bbo
\label{af1000}
\sqrt{J/2}\; \Phi'(X,r)= \chi(X)\phi(r).
 \ebo      
The action with the factorized field becomes,
\bbo
\label{af2}
 \backo S= M^4
\!\!\int d^4X d^4r\; 
\bl Z|\phi(r)|^2|\partial_X\chi(X)|^2 
+Z|\chi(X)|^2|\partial_r\phi(r)|^2 
+2{g_0^2 N_c}D_F(2r)|\chi(X)\phi(r) |^2 
\br,
\ebo      
A canonical normalization of the $\chi(X)$ kinetic term  (which is the requirement
of a normalized Noether current, such as 
 $i\chi^\dagger \frac{\overleftrightarrow\partial}{\partial X^\mu} \chi$ 
 \cite{main})
dictates a normalization constraint on $\phi(r)$. Define the Lorentz invariant normalization:
  \vspace{-0.0in}
\bbo \label{basenorm}
 1= ZM^4\!\!\int d^4 r\; |\phi(r)|^2 .
\ebo
Following the elementary two-body kinematics, where the relative time
disappears in the rest frame, then  $\phi(r)\rightarrow \phi(\vvr)$ becomes a static field that
has no dependence upon $r^0$  (this can be formally
handled with Lorentz invariant constraints as in \cite{main}).
We can then  define $Z$,  
 \bbo 
 \label{phinorm00}
 1 = ZM \!\int\! dr^0 \equiv ZMT= \epsilon ZM_0 T ,
   \ebo   
 which, in turn, dictates a normalization for $\phi(\vvr)$:
  \bbo 
 \label{phinorm}
  1=  M^3 \int d^3 r\; |\phi(\vec{r})|^2.  
   \ebo  
The condition $1= \epsilon Z M_0 T$ 
removes the relative time, $T=\int dr^0$, from the kinetic terms.
Note that $\phi(\vvr)$ is then  dimensionless with eq.(\ref{phinorm}).
 The action becomes,
  \bbo
  \label{intD}
\backo S =
\!\!\int d^4X \bl
 |\partial_X\chi(X)|^2 
+  
|\chi(X)|^2 M^3\!\!\int \!\! d^3r\; \left(
-|\partial_{\vvr}\phi(\vvr)|^2 
+\int dr^0  \;{2g_0^2 N_c M }D_F(2r^\mu)|\phi(\vvr)\; |^2
\right)\br.
 \ebo 
(Note $|\partial_r\phi|^2= |\partial_{r^0}\phi|^2 - |\partial_{\vec{r}}\phi|^2$).
We finally integrate over $r^0$ in the interaction term:
       \bbo\label{Ypot000}
\backo\!\!\!\int\!\! dr^0  D_F(2r)
= -\!\!\int\!\!  dr^0 \frac{d^4q}{(2\pi)^4}
\frac{1}{q^2-M_0^2}e^{2i q_\mu r^\mu }
= \half \int\!\!   \frac{d^3q}{(2\pi)^3}
\frac{1}{\vec{q}{\;}^2+M_0^2}e^{2i q_\mu r^\mu }= -\half V_0(2|\vvr|).
\ebo 
The $\vec{q}$ momentum integral yields the familiar Yukawa potential
(where $2r $ is the separation of the particles), 
\bbo
\label{Ypot}
 V_0(2r) =-\frac{ e^{-2M_0 |\vvr|}}{8\pi |\vvr|}.
 \ebo      
The action then becomes,
  \bbo
  \label{intDf}
\backo S =
\!\!\int d^4X \bl
 |\partial_X\chi(X)|^2 
+  
|\chi(X)|^2 M^3\!\! \int \!\! d^3r\; \left(-
|\partial_{\vvr}\phi(\vvr)|^2 
+{g_0^2 N_c  M}\frac{ e^{-2M_0 |\vvr|}}{8\pi |\vvr|}|\phi(\vvr) |^2
\right) \br.
 \ebo 
 Note that, in the limit of suppressing the
$\vec{q}{\;}^2$ in the denominators of the integrands of eq.(\ref{Ypot000}), we obtain the large $M_0^2$ limit
of the potential
(using $J=2^4$, 
and $\delta^3(\vec{r})=(4\pi r^2)^{-1}\delta(r)$):
\bbo
\label{Ypot2}
V_0(2r)\rightarrow -\frac{1}{M^2_0}\delta^3(2\vvr)=-\frac{2}{JM^2_0}\delta^3(\vvr)=-\frac{1}{2\pi J M^2_0 r^2}\delta(r).
 \ebo      
 We emphasize that the theory remains Lorentz invariant, albeit not manifestly so.
Note that it is convenient to write things as,
in the format: 
\bbo
\label{51}
\backo
S = 
 \int\!\! d^4X\; \bl |\partial_{X}\chi |^2 - |\chi |^2{\cal{M}}^2
\br
\nonumbo \backo
\mu^2 = {\cal{M}}^2
\equiv M^3\!\! \int \!\! d^3r\; \bl |\partial_{\vvr} \phi |^2-g_0^2N_c 
M\frac{ e^{-2M_0 |\vvr|}}{8\pi |\vvr|}|\phi|^2\br,
  \ebo      
  where $\mu^2$ is the eigenvalue of the ``Hamiltonian,'' ${\cal{M}}^2$.
  We see that the action of $\phi$ is ``nested'' within the action
  of $\chi$, where $\chi$ controls the motion of
  the collective state and $\phi $ describes the internal relative motion of the constituents.
${\cal{M}}^2$ is a Hamiltonian for the static
internal wave-function $\phi(\vec{r})$.

Extremalizing ${\cal{M}}^2$ with respect to $\phi(r)$ 
implies the Schr\"odinger--Klein--Gordon (SKG) equation for
an s-wave ground state and its eigenvalue,
 $\mu^2$:
\bbo
\label{SKG}
\backo\!\!\!\!
-\bl\frac{\partial^2 }{\partial r^2}+\frac{2}{r}\frac{\partial }{\partial r}\br\phi(r)-g_0^2N_c M\frac{ e^{-2M_0 r}}{8\pi r}\phi(r) =\mu^2\phi(r)
   \ebo      
 We then see that ${\cal{M}}^2=\mu^2$ is then the physical mass  of the bound state. The $\chi$ action in
 any frame is manifestly Lorentz invariant (where we include a quartic term, which
 will be developed below):
        \bbo
\backo\!\!\!
S = \!\!
 \int\!\! d^4X \bl |\partial_{X}\chi(X) |^2 - \mu^2|\chi(X) |^2
-\frac{{\lambda}}{2}|\chi(X)|^4\br.
   \ebo      
 The Yukawa potential has a critical coupling, $g_0=g_{c}$, 
 where the eigenvalue is then $\mu =0$.
 For $g_0>g_{c}$ then $\mu^2<0$, and we have spontaneous symmetry breaking.

\subsection{A Simple example:}

As an aside we note that we can represent a two body open scattering state as a bilocal
wave-function with $\phi(\vvr)=N\exp(2i\vec{Q}\cdot\vvr)$.
$\phi(r)$ is then ``non-normalizable'' (requiring box normalization)
and is then  $M^3N^2\int d^3r|\phi(\vvr) |^2=1$. In
the center of mass rest frame action becomes,
\bbo
  \label{af3}
\backo \! S_K\!=\!
V_3\int\! dX^0\; \bl |\partial_0\chi(X^0)|^2 - 4\vec{Q}\;{}^2 |\chi(X^0)|^2\br
\ebo 
 This is a state described by $\chi(X)$ 
 which 
 satisfies the equation of motion,
\bbo    
 \partial_0^2\chi + \mu^2\chi =0   \qquad \mu^2=4\vec{Q}\;{}^2,
   \ebo 
This is a
zero 3-momentum two body
scattering state of invariant mass $2|\vec{Q}|=\mu $, with conventional volume
normalization $\sim V^{-3/2}$.  Technically, the experimentalists' ``invariant mass'' of a two body state is not a mass at all; a mass appears in the trace of the stress tensor and massless particles
have a vanishing trace. In theories like QCD the mass scale is set by the ``trace anomaly,''
and presumably the scale, $M_0$, would emerge in  similar fashion in the coloron theory.

 \subsection{The Induced Bound State Yukawa Interaction}
 
The Yukawa interaction of the bound state with
the free scattering state fermions is now induced from the second term, $S_Y'$,  in eq.(\ref{5NJL}). We have,
noting  eqs.(\ref{5NJL}, \ref{af1000} ):
\bbo 
\label{5NJL000}
\backo
S_Y' =
g^2_0\sqrt{N_c}M^2\!\!\int\!\! d^4x d^4y \;[\bar{\psi}_L(x)\psi_{R}(y)]_{f}  D_F(x-y)\;\Phi^0(x,y) {+h.c.}
\nonumbo
=  \sqrt{2N_cJ} g^2_0 \epsilon^2 M_0^2
\int\!\! d^4X d^4r \;[\bar{\psi}_L(X\!+\!r)\psi_{R}(X\!-\!r)]_{f}D_F(2r)\;\chi(X)\phi(\vvr) {+h.c.}.
\ebo
Consider the pointlike limit of the potential,
eq.(\ref{4NJL}),  $ D_F(2r)\rightarrow (JM_0^2)^{-1} \delta^4(r)$:
\bbo
\backo
S_Y'
\rightarrow
\sqrt{2N_c/J} g^2_0 \epsilon^2
\int\!\! d^4X\; [\bar{\psi}_L(X)\psi_{R}(X)]_{f}\chi(X)\phi(0) {+h.c.}.
\ebo
We therefore see that the induced Yukawa coupling to the field $\chi(x)$
in the pointlike limit {\em of the potential} (which should be a reasonable low energy
approximation) is:
\bbo
\label{gy}
g_Y= \hat{g}_Y\phi(0)\qquad\makebox{where,}\qquad   \hat{g}_Y\equiv g_0^2\epsilon^2\sqrt{2N_c/J} 
=g_0^2\epsilon^2\sqrt{3/8}.
\ebo
We emphasize that this is a significant result and fundamentally different than the
NJL model result.  We have taken the pointlike limit of the potential as in the
NJL model, but obtain a result that is dependent
crucially upon the non-pointlike internal wave-function $\propto \phi(0)$. The implication
 is that a strong coupling, $g_0^2$,  can produce, in principle, a small
Yukawa coupling if $\phi(0)<\!\!< 1$. 
In the usual pointlike NJL model the induced Yukawa coupling runs  to smaller
values in the IR, but it does so only logarithmically, via the RG.  Here the behavior
of $\phi(0)$ is a suppression of $g_Y$  that will be seen, in the next section,
to be power-law $\sim \sqrt{|\mu|/M_0}$ for small $\mu$,  near the critical coupling.

\section{The Schr\"odinger-Klein-Gordon (SKG) Equation}

While formally similar to the non-relativisitic Schr\"odinger equation,
 the SKG
equation, eq.(\ref{SKG}), has key physical differences:
\begin{itemize}
\item the potential has dimension (mass)$^2$, rather than energy;
\item the eigenvalue describes resonances for positive $\mu^2$; 
\item a negative eigenvalue, $-|\mu|^2$, implies  vacuum instability
and spontaneous symmetry breaking. 
\end{itemize}

Mainly the Hamiltonian, ${\cal{M}}^2$ of eq.(\ref{51}), which generates the SKG equation,
is amenable to variational calculations as we show below. 
We presently give some examples of solutions and stress
some subtleties, though much can be done to refine and extend this discussion.
The solutions allow the computation of the induced top quark Yukawa coupling
of the bound state to free fermions, $g_Y$, via the wave-function
at the origin, $\phi(0)$, from which one can extract $M_0$, the mass scale
of the potential (i.e. the coloron mass).  

 A negative eigenvalue of the Schr\"odinger equation defines our conventional
 view of a non-relativistic bound state.  However, in the relativistic case, 
 for a pair of chiral fermions, the SKG equation with a bound state solution
 implies a negative $\mu^2$.   
 This is, of course, the
 behavior of $\Sigma$-model in QCD and the BEH boson in the standard model
 and requires
 additional physics to stabilize the vacuum, such as quartic interactions. 
 Hence, the general result is that 
 a scalar bound state of massless
 chiral fermions in the symmetric (unbroken) phase must 
 either be an unstable resonance (subcritical coupling and positive $\mu^2$), which decays rapidly to its
 constituents,
 or tachyonic (supercritical coupling, negative $\mu^2$) leading to a chiral instability of the vacuum.

\subsection{Variational Calculation Determining  $\epsilon$ \label{variational}}

 We defined $M=\epsilon M_0$, as the mass scale in the ansatz, 
introducing the parameter $\epsilon$.  
The largest mass scale at which the static potential approximation is applicable
is $M=M_0$, hence $\epsilon \leq 1$.
$\epsilon $ is seen to multiply the underlying coupling constant, $\tilde{g}_0^2 = \epsilon g_0^2$.
The largest value of  $\tilde{g}_0^2$,
is therefore $g_0^2$, hence $\epsilon=1$ implies the smallest
possible critical value of the underlying coupling $g_0^2$.
We view $\epsilon$ as part of the wave-function ansatz, and allow the variational
calculation of the bound state mass to determine $\epsilon$ by minimization of the Hamiltonian.
By ``Hamiltonian'' we mean ${\cal{M}}^2$ of eq.(\ref{51}).

 A solution to the SKG equation for the eigenvalue
 can be approximated by a variational calculation. 
  For the present  calculation we assume
 an ansatz consisting of a Hydrogenic 
 wave-function, $\widetilde{\phi}(r) = Ae^{-Mr}$, with $M=\epsilon M_0$ and $\epsilon$ as the variational
 parameter, and $M_0$ is the scale of the Yukawa potential.
 This cannot be a precise description near criticality where the eigenvalue $\mu^2$ is small because
 it lacks the large distance tail  $\propto e^{-|\mu| r}/r$ for small $\mu$, 
 however, it conveniently illustrates how $\epsilon\rightarrow 1$ is established dynamically for the bound state.
 
The normalization condition for the ansatz is defined in  eq.(\ref{phinorm}),
  \bbo
  \backo
 1= 4\pi A^2\epsilon^3 M_0^3\int_0^{\infty}\!\!\!\! e^{-2\epsilon M_0r}r^2 dr; \qquad
 \makebox{hence,} \qquad A^2=\frac{1}{\pi }.
 \ebo\normalsize
The Hamiltonian ${\cal{M}}^2$  and eigenvalue $\mu^2$ of eq.(\ref{51}) with $M=\epsilon M_0$ is therefore,
 \bbo
\label{510}
{\cal{M}}^2=\epsilon^3M_0^3\!\!\int\!\!d^3r\; \bl |\partial_{\vec{r}}\phi |^2+g_0^2N_c \epsilon M_0 V_0(2r)|\phi|^2\br \equiv \mu^2,
\ebo
where, 
 $V_0(2r)=- e^{-2M_0 r}/{8\pi r}$,
as in eq.(\ref{Ypot}), with the fixed coloron mass $M_0$ (no $\epsilon$ factor is present in $V_0(2r)$).
 In what follows we will use a definition
of the coupling, 
\bbo
\kappa =\frac{g_0^2N_c}{4\pi}\qquad \kappa_{c_{NJL}}= 2\pi,
\ebo
(where for reference, we quote the implied NJL critical value $\kappa_{c_{NJL}}$).
 We then compute the eigenvalue $\mu^2 = {\cal{M}}^2$ as a function
 of $\epsilon$ and $\kappa$:
 \bbo
 \label{epsvar}
 {\cal{M}}^2=A^2  \epsilon^3 M_0^3
 \int d^3r \bl |\partial_{r}e^{-\epsilon M_0r} |^2
 -\frac{\kappa\epsilon M_0}{2} \frac{e^{-2M_0r}}{r} |e^{-\epsilon M_0r}|^2\br 
 \nonumbo
=M_0^2\bl \epsilon^2 - \frac{\kappa \epsilon^4 }{2(1+\epsilon)^2}\br.\;\;\;   
\ebo
  
{
\begin{figure}
	\centering
	\includegraphics[width=0.6\textwidth]{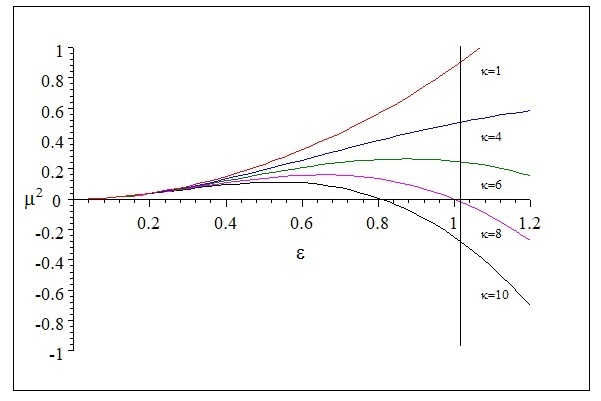}
	\caption{\footnotesize\normalfont ${\cal{M}}^2=\mu^2$  of eq.(\ref{510}) is plotted vs. $\epsilon$, with $M_0=1$
	for values of $\kappa=g_0^2 N_c/4\pi = (1,4,6,8,10)$. 
	The critical  coupling is the value of $\kappa$ for which a massless bound state first
    occurs, i.e., ${\cal{M}}^2=\mu^2=0$. This happens for the $\kappa=8$ curve, 
    which intersects  
    the $\epsilon =1$ vertical line and has $\mu^2=0$. 
    Hence $\kappa_c=8$ is the critical coupling with this ansatz. }
	\label{fig1}
\end{figure}
}

In Fig.(\ref{fig1}) we plot a family of curves of ${\cal{M}}^2=\mu^2$ for various values of  $\kappa$ 
as function of $\epsilon$.
We see  that the extremal (smallest) value for positive $\mu^2$ corresponds to
$\mu^2=0$ and occurs for any $\kappa< 8$. On the other hand,
for $\kappa> 8$ (purple curve) the extremal,
most negative value of $\mu^2={\cal{M}}^2 $,
occurs when $\epsilon\rightarrow 1$.  Hence, the critical coupling for this ansatz is $\kappa_c=8$ and we then have,
 \bbo  \frac{\kappa_c}{2\pi} =\frac{ g^2_{c}N_c}{8\pi^2}=\frac{4}{\pi}= 1.27,
 \ebo 
 compared to the NJL critical value $1.00$ (reflecting the crudeness of the ansatz).
 
However, we see  an interesting result:  We find that $\epsilon=1 $ is the true minimum
for any $g_0^2>g_c^2$ where $g_c^2=4\pi\kappa_c/N_c$ is the critical coloron coupling. In this case $\mu^2<0$ and
we will have spontaneous symmetry breaking. 
For $g_0^2< g_c^2$, then $\mu^2>0$, and there is a true minimum at $\epsilon=0$. However, 
for subcritical couplings $g_0^2< g_c^2$, in the range $8\gta \kappa \gta 6$, we see
from the Fig.(\ref{fig1}) that there are a quasi-stable minima at $\epsilon=1$,
with $\mu^2>0$.
Here we expect to have resonances
which decay to free fermion pairs.
We believe the quasi-stable minimum reflects the presence of such resonances, 
where the wave-function with $\epsilon\sim 1$ would 
tunnel through the $\epsilon <1$  barrier to reach the true minimum $\epsilon =0$.   As $g_0^2<\!\!< g_c^2$,
roughly $\kappa \lta 6$, the quasi-stable minimum
disappears, and the true minimum 
is at $\epsilon =0$, and no bound state resonances forms.
 
The variational result for $g^2_c$ with this ansatz 
gives a false value for the  normalized trial wave-function 
 at the origin for critical coupling,
$\phi(0)=  {1}/{\sqrt{\pi} }$.
The reason is, of course, that the ansatz does not include
a $e^{-|\mu|r}/r$ tail at large $r$, which 
 significantly affects the normalized $\phi(0)$.
Below we do refined calculations that demonstrate the effects of the large distance
tail of $\phi(r)$.

\subsection{Exact Criticality of the Yukawa Potential \label{exact2}}

The coloron model  furnishes a  direct UV
completion of the NJL model.  It leads to an SKG potential of the Yukawa form
which has a critical coupling, $g^2_0=g^2_{c}$.
 The critical coupling is that value of $g_0^2$
for which the eigenvalue $\mu^2$ is zero.
We wish to determine $g_{c}^2$ exactly. 

The criticality of the Yukawa potential in the non-relativistic
Schr\"odinger equation is widely discussed in the literature in the context of ``screening''
(see \cite{Edwards} and references therein).
The non-relativistic Schr\"odinger equation $r=|\vvr|$ is:
 \smbbo
 \label{SKG01}
-\nabla^2\psi - 2m_e\alpha\frac{e^{-\mu r}}{r}\psi=2m_eE,
 \smebo
 with $m_e$ the electron mass, 
and eigenvalue $E=0$ occurs for a critical screening
with $\mu=\mu_c$. A numerical analysis yields,
\cite{Edwards},
 \smbbo
 \label{Edwards2}
\mu_c= 1.19061 \;\alpha m_e.
 \smebo
For the spherical SKG equation in the coloron model eq.(\ref{SKG}) we have from the Hamiltonian,
  \bbo
\label{SKG00}
\backo
-\nabla^2\phi(r)-g_0^2N_c M_0\frac{ e^{-2M_0 |\vvr|}}{8\pi |\vvr|}\phi(r) =\mu^2\phi(r),
 \ebo\normalsize
where we assume  $\epsilon=1$ as determined by the variational calculation above.
 
 We can obtain the critical coloron model coupling constant by 
comparing, eq.(\ref{SKG01}) and eq.(\ref{SKG00}). We have,
\bbo
2m_e\alpha \rightarrow  g_0^2 N_c M_0/8\pi,
\qquad \qquad 
\mu_c\rightarrow 2M_0,
\ebo
then
substituting into eq.(\ref{Edwards2}),  
$ 2M_0= 1.19061 (g_0^2 N_c M_0/16\pi)$, and therefore,
\bbo
\label{exact}
\left.\frac{g_0^2N_c}{8\pi^2}\right|_c= \frac{4}{(1.19061)\pi} =1.06940.
 \smebo
By comparison, the loop level NJL critical value of eq.(\ref{0NJL4}) is, 
\bbo
\label{critc2}
\left.\frac{g_c^2N_c}{8\pi^2}\right|_{NJLc}=1.00.
\ebo
Hence, we see that the NJL quantum critical coupling has a remarkably
similar numerical value to the classical critical coupling.
(It is beyond the scope of the present paper to understand why these
are not identically equal!)

\subsection{Spherical Potential Well \label{potwell}}

To compute the induced Yukawa coupling, $g_Y$,  we then need to calculate $\phi(0)$.
For a convenient solvable potential problem and warm-up exercise, we turn to  
the spherical potential well,
\bbo
\frac{e^{-2M_{0}r}}{8\pi r}\longrightarrow \frac{\lambda }{%
8\pi }M_{0}\theta (1-M_{0}r).
\ebo
We match the integral over the spherical potential well to the integral over
the Yukawa potential, which determines  $\lambda=3/4$.
Once matched, the spherical well
can be used as an approximation to the Yukawa potential. Moreover, the formula
$ g_{Y}=g_{0}^{2}\sqrt{2N/J}\;\langle \phi(0) \rangle $ 
of eq.(\ref{gy}) is modified by replacing $\phi(0)$ by it's volume average 
$\langle \phi(0) \rangle$
over the well, $r< M_0^{-1}$.
The SKG equation for the potential well is:
\bbo
-\bl \frac{\partial ^{2}}{\partial r^{2}} +\frac{2 }{r}\frac{\partial }{\partial r} \br \phi(r)-  \frac{3g_0^{2}NM^2_{0}}{32\pi }\theta (1-M_{0}r)\phi(r)=\mu ^{2}\phi(r),
\ebo
with  solution:
\bbo 
\label{soln}
 \phi(r) =\phi_1(r) \theta(1-M_{0}r)+ \phi_2(r)\theta(M_{0}r-1) \qquad \makebox{where,}\qquad
\nonumbo \phi_1(r) = \frac{A\sin (kr)}{r}  \qquad   \phi_2(r) = \frac{Be^{-\left| \mu\right|(r-M_{0}^{-1})}}{r}.
\ebo
Continuity  at $M_{0}r=1$ then implies,
\bbo
A\sin (kM_{0}^{-1})=B;\;\;\;\qquad Ak\cos (kM_{0}^{-1})=-B\left| \mu \right|.
\ebo
Critical coupling implies $\left| \mu \right| =0$, hence,
$\cos (kM_{0}^{-1})=0$, which yields $k=\frac{\pi}{2}M_0$ and $ A=-B$.
Therefore, we obtain the critical coupling for the spherical well,
\bbo 
\label{swcc}
\left( \frac{\pi }{2}M_{0}\right)^{2}-
\frac{3g_c^{2}N_c}{32\pi 
}M_{0}^{2}=0,
 \qquad\makebox{hence,}\qquad 
\frac{g_{c }^{2}N_c}{8\pi ^{2}}=\left( \frac{\pi }{3}\right)
= 1.0472.
\ebo
This is 
close to the NJL result $(=1.0)$, or to Yukawa $(=1.06940)$.
Note that at critical coupling
we expect resonances at $\mu_r\approx \frac{\pi}{2}\sqrt{N^2+2N}M_0$, for $N=1,2,..$.

The normalization of the ansatz of eq.(\ref{phinorm}) determines the coefficient, $A$,
and is dominated by the tail of $\phi(r)$ for small $\mu $:
\bbo
1= 4\pi A^{2}M_{0}^{3}\int r^{2}dr\left| \phi(r) \right|^{2}\approx 
4\pi A^{2}M_{0}^{3}\int_{0}^{\infty }\frac{e^{-2\left|\mu \right| r}}{r^2} r^2 dr
=\allowbreak \frac{2\pi A^{2}}{\left| \mu \right| }(M_{0})^{3}\nonumber
\nonumbo
A\approx \frac{1}{M_{0}}\left( \frac{\mu}{2\pi M_0 }\right)^{1/2}=\allowbreak \frac{%
0.398\,94}{M_{0}}\left(\frac{\mu}{M_0 }\right)^{\!1/2}.
\ebo
The wave-function at the origin is technically given by,
\bbo
\label{techphi}
\phi(0) = \left( \frac{A\sin (kr)}{r}\right)_{r\rightarrow 0}=Ak= \frac{\pi}{2}AM_0
=\left( \frac{\pi \mu}{8 M_0 }\right)^{\!\!\!1/2}.
\ebo
We see that $\phi(0)$  is therefore suppressed as $\sim \left( |\mu|/M_0 \right) ^{1/2}$.
However, low momentum fermions in the Yukawa interaction  would experience the volume average of the wave-function in the well,
(recall above  $k=\pi M_0/2$):
\bbo
\langle\phi(0)\rangle = 4\pi {\cal{N}}^{-1}\int_0^{M^{-1}_0}\!\! \left( \frac{A\sin (kr)}{r}\right) r^2 dr
\;\;\;\makebox{where,} \;\;\;{\cal{N}} =\frac{4}{3}\pi M_0^{-3}
\nonumbo
\makebox{hence,}\;\;\; \langle\phi(0)\rangle =\frac{6\sqrt{2}}{\pi^{5/2}}\sqrt{|\mu|/M_0}.
\ebo
We can then compute $M_0$ from an input $|\mu|$ and the value of the Yukawa coupling using eq.(\ref{gy}):
\bbo
1\approx g_{Y}= g_{c}^{2}\sqrt{2N_c/J}\langle \phi (0)\rangle = \left( \frac{\pi }{3}%
\right) \frac{8\pi ^{2}}{N_c}\sqrt{3/8} \langle\phi(0)\rangle \approx 8.1867 \sqrt{|\mu|/M_0}.
\ebo
Applying this to top quark condensation we input the (symmetric phase) BEH boson Lagrangian mass, $|\mu|=88$ GeV, to obtain:
\bbo
M_0 \approx 5.9 \;\;\makebox{TeV}.
\ebo  
The ``skeletal solution'' of Section \ref{skeletal} below yields a similar $M_0\sim 6$  TeV result.
Note that if we had used $\phi(0)$, in the potential well, which is much larger than the average
$\langle\phi(0)\rangle$ as given by eq.(\ref{techphi}), we would obtain $M_0 \approx  10$ TeV.

\subsection{Significantly Reduced Fine Tuning Due to Dilution  \label{finetune}}

The spherical potential well also illustrates a key feature of the fine--tuning and the nature
of the phase transition.  If we consider the Hamiltonian
we have:
\bbo
\label{hamwell}
\mu^2= 
4\pi M_0^3\bl\int_0^{M_0^{-1}} r^2 dr\left((\partial_r\phi_1)^2 -  \frac{3g_0^{2}NM^2_{0}}{32\pi }\phi_1^2\right)
+ \int_{M_0^{-1}}^\infty r^2 dr(\partial_r\phi_2)^2 \br 
\nonumbo
=4\pi M_0^3A^2\bl\int_0^{M_0^{-1}} r^2 dr\left(\frac{\pi^2 }{4}M^2_{0}-\frac{3N_cg_c^{2}}{32\pi }M_{0}^{2} \right)\frac{\sin^2 (kr)}{r^2}
+ \int_{M_0^{-1}}^\infty r^2 dr \mu^2 \frac{e^{-\left| 2\mu\right|(r-M_{0}^{-1})}}{r^2} \br 
\nonumbo
=|\mu| \bl M_0\left(\frac{\pi^2 }{4}-\frac{3N_cg_c^{2}}{32\pi } \right)
+ |\mu|  \br .
\ebo
Here we see that the critical behavior is significantly modified with respect to the NJL model.
In the NJL model the critical behavior in $g_0^2$ near its critical value $g_c^2$
is analogous to that of
a second order phase transition,
\bbo
\mu^2 \sim M_0^2\left(1 - \frac{g_0^2}{g_{c}^2}\right),
   \ebo
which  implies significant fine--tuning to obtain a large hierarchy, 
\bbo
\frac{\delta  g_0^2}{g_{c}^2} \sim \frac{\mu^2}{M_0^2}.
   \ebo

However, in the present framework we see from eq.(\ref{hamwell}) that, {\em for a negative $\mu^2$ },
the large distance tail and its
dilution effect (where $\phi(0)\propto \sqrt{|\mu|/M_0}$) modify this relationship as:
\bbo
\mu^2 \sim \phi(0)^2\left(M_0^2 - \frac{g_0^2}{g_{c}^2}M_0^2\right)
\sim  \frac{|\mu|}{M_0} \left(M_0^2 - \frac{g_0^2}{g_{c}^2}M_0^2\right)
\nonumbo
\rightarrow \frac{\mu^2}{|\mu|} \sim  {M_0}\left(1 - \frac{g_0^2}{g_{c}^2}\right).
   \ebo
We thus obtain a linear relationship between $\mu$ and $M_0$ near criticality,
\bbo
\frac{\delta  g_0^2}{g_{c}^2} \sim \frac{|\mu|}{M_0}.
   \ebo 
Due to the dilution effect we can tolerate
  significant departures from criticality $g_c^2 + {\cal{O}}\left( \mu/M_0 \right)$
  and we will still have an expectation  value of Hamiltonian of $\sim {\cal{O}}(\mu^2)$.
  The specification of the critical coupling is therefore made much less stringent by the 
  large distance tail and dilution effect of the wave-function.
  
 This result can be checked by perturbing the potential well solution around the critical coupling value.
 We have for $g_0^2 = g_c^2 +\delta g^2$ and $k\rightarrow k+ \delta k$,
\bbo
A\sin ((k+ \delta k)M_{0}^{-1})=B;\;\;\;\qquad A(k+\delta k)\cos ((k+ \delta k)M_{0}^{-1})=-B\left| \mu \right|.
\ebo
Expanding in $\delta k $ and using $kM_{0}^{-1}=\pi/2$, 
\bbo
B= A;\qquad  B= A\frac{k\delta k}{M_0|\mu|} \qquad \makebox{hence,} \qquad    \delta k= \frac{ 2|\mu|}{\pi} .
\ebo
The potential well yields,:
\bbo 
\left( \frac{\pi }{2}M_{0}+\delta k\right)^{2}-
\frac{3N_c(g_c^{2}+\delta g^2)}{32\pi 
}M_{0}^{2}=\mu^2
 \qquad\makebox{hence,}\qquad
\frac{\pi \delta k}{M_0} -\frac{3N_c( \delta g^2)}{32\pi }={\cal{O}}(\mu^2/M_0^2)\approx 0,
\qquad 
\nonumbo 
\ebo
so we obtain  (using  eq.( \ref{swcc})):
\bbo 
\frac{\delta g^2}{g_c^2} =  \frac{4\delta k}{\pi M_0}=\frac{8}{\pi^2}\frac{|\mu|}{ M_0} + {\cal{O}}(\mu^2/M_0^2),
\ebo
confirming the linear relation between ${\delta g^2}/{g_c^2}$ and $|\mu|$.

We have illustrated the bound state with the spherical well potential since
it is simple and can be readily verified.
Note that, if we had used the NJL criterion for fine--tuning,  we would have ${\delta  g_0^2}/{g_{c}^2} \sim {\cal{O}}(10^{-4})$.
The linear relationship yields tuning at the reduced few $\%$ level,  due to
dilution from the tail of the internal wave-function.

 \subsection{Skeletal Solution of the SKG Equation \label{skeletal} }

 We can give a very simple illustrative result that faithfully reproduces the mass scale $M_0$
 as determined from low energy inputs.
 Here we assume the short distance potential is given by the $\delta-$function limit
 of the Yukawa potential, and $\phi(0)$. 
 
 We can approximate the Yukawa potential by a $\delta$-function at short distance
 using eq.\ref{Ypot2} (recall $J=16$ is the Jacobian passing from $(x,y)$ to $(X,r)$):
 \bbo
\label{}
\frac{ e^{-2M_0 |\vvr|}}{8\pi |\vvr|} \rightarrow \frac{1}{M^2_0}\delta^3(2\vvr)=\frac{2}{JM^2_0}\delta^3(\vvr).
 \ebo
 Consider
 the form of the potential that appears in the coloron model,
 \bbo
 -\nabla^2 \phi -g_0^2N_c M_0\frac{ e^{-2M_0 |\vvr|}}{8\pi |\vvr|}\phi(r)\longrightarrow -\nabla^2 \phi 
 -\frac{2g_0^2N_c }{JM_0}\delta^3(\vec{r})\phi(0) =\mu^2\phi
 \ebo
 Using $\nabla^2(1/r)=-4\pi \delta^3(\vec{r})$,
the large distance solution is then,
 \bbo
 \phi(r) = \frac{c e^{-|\mu| r}}{r}; \;\;\;\;\;\makebox{where,}\;\;\;\;\;
 c=\frac{g_0^2N_c }{2\pi JM_0}\phi(0).
 \ebo
 At critical coupling $g_0^2=g^2_{c}$, and then  $\mu^2=\eta^2 \approx 0$, where $\eta$ is an infinitesimal 
 infrared cut-off mass scale that is necessary to give $\phi(r)$ a finite normalization:
\bbo
 1= M_0^3 4\pi \! \int_0^\infty \! r^2 dr\; \frac{c^2}{r^2}e^{-2|\eta| r}
 = 4\pi \frac{M_0^3}{2|\eta|}  \left(\frac{g_c^2N_c}{2\pi JM_0}\phi(0)\right)^{\!\!2},
\ebo
 hence, with $g_c^2N/8\pi^2=1$,
 \bbo
\phi(0)=\frac{2\sqrt{2}}{\pi^{3/2}} \sqrt{ \frac{|\eta|}{M_0}} \sim 0.
 \ebo
 and we see that  $\phi(0)$ is diluted to zero in the critical coupling case as $\sim \sqrt{\eta}\rightarrow 0$.
 
 If the coupling is chosen to be supercritical, $g_0^2=g_c^2 +\delta g_0^2$,
 then the mass $\mu^2$ is physical and the normalization becomes,
\bbo
\phi(0)=\frac{2\sqrt{2}}{\pi^{3/2}} \sqrt{ \frac{|\mu|}{M_0}} =0.50795\sqrt{ \frac{|\mu|}{M_0}}.
 \ebo
 If we 
 assume we are close to critical coupling, $g_0^2N_c\approx 8\pi^2$, 
 we can approximate the result for the Yukawa coupling and use as input $|\mu|= 88$ GeV: 
 \bbo
 \backo\backo\backo
 1\approx g_Y= \frac{8\pi^2}{3}\sqrt{\frac{2N_c}{J}}\;\phi(0)=16.12\;\phi(0)
 \approx 8.187 \sqrt{\frac{|\mu|}{M_0}};\;\;\;\;\;\makebox{therefore,}\;\;\;\;
 M_0 = 5.9\; \makebox{TeV.}
 \ebo
 The skeletal model tells us nothing about what determines the critical coupling, $g^2_c$.
 This is determined by the short distance solution inside the potential
 and its matching onto the large distance solution as we saw in the previous 
 spherical well example. However, it is sufficient to assume the critical behavior
 which leads to  the approximate scale invariance (small $|\mu|$) of the large distance solution 
 to obtain the relation between $g_Y$ and $M_0$.

  \subsection{Numerical Integration the SKG Equation \label{numerical}}
  
     Central to this theory is determining the mass $M_0$ given the Yukawa coupling of the
  top quark, $g_Y\equiv g_{top}\approx 1$ which, via the wave-function $\phi(r)$,
  determines $|\mu|/M_0$, where $|\mu|$ is the symmetric phase BEH mass, $|\mu| \sim 88$ GeV.
  In  preceding estimates we considered the spherical well solution, or we took the short-distance ($\delta-$function) limit of the potential,
  while allowing the wave-function spread as $\phi(r)\sim e^{-|\mu|}/r$. We then
  related $g_Y$ to $\phi(0)\propto \sqrt{|\mu|/M_0}$.  However, we can obtain a complete 
  numerical solution of the extended wave-function to the SKG equation with the Yukawa potential, We can then
   relate $g_Y$ to $\phi(r)$ integrated over the potential, without taking the $\delta-$function limit.
   
  The SKG equation  eq.(\ref{SKG}) with $M_0=1$  takes  the form in $u(r)$ where $u(r)=r\phi(r)$:
\bbo
\label{uu}
u''(r) +\frac{g^2N_c}{8\pi r}e^{-2r}u(r)=-\mu^2 u(r)
\ebo
It is easy to obtain reasonable approximate numerical solutions, following
 ref.\cite{Edwards}. We use numerical inputs  $g_c^2N_c/8\pi=\pi f =\pi \times 1.06940\approx 3.36$,
 where $g_c^2N_c/8\pi^2=1.06940$ is the exact
critical coupling of eq.(\ref{exact}).
 The critical coupling corresponds to the solution $|\mu|=0$, 
for which $u(r)\rightarrow$(constant), as $r\rightarrow \infty$.  We find numerically in
the critical case that $u(r)\rightarrow 0.3240$ as $r\rightarrow \infty$.
The numerical solutions are shown in Fig.(\ref{figM3}).
The resulting  numerical solutions can be approximately fit by the function $u_{fit}(r)$,
\bbo
\label{form}
u_{fit}(r) = 0.3240\bl \bl 1-\left(1-\frac{r}{R_0}\right)^{\!p} \;\br \theta (R_0-r)+e^{-|\mu|(r-R_0)}\;\theta(r-R_0)\br
\ebo
where $R_0=1.8(M_0)^{-1}$ and  $\mu=0$ in the exact critical case.
The best  ``visual fit'' is obtained for $p=4$
as shown in  in Fig.(\ref{figsolnsfit}), curve (b).

        {	\hspace{2.0in}
\begin{figure}
	\includegraphics[width=1.0\textwidth]{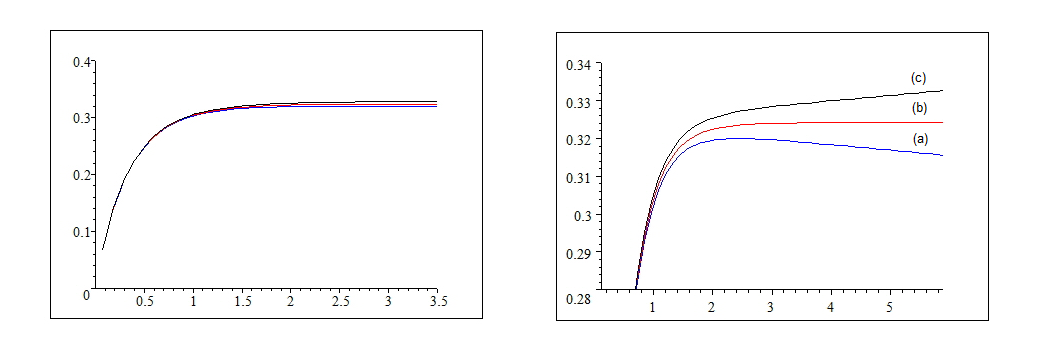}
	\vspace{-0.2in}
	\caption{\footnotesize \normalfont Numerical solutions to eq.(\ref{uu}) for 
	for ${g_0^2N_c}/{8\pi}=c$, where:  
	(a) (blue) $c=3.37$  (b) (red) $c=3.36$;  (c) (black) $c=3.35$. We see that the curve (b) flattens, corresponding to 
	$|\mu|=0$ and $\phi(r)\sim u(\infty)/r$ critical coupling, with $u(\infty)=0.3240$, (see ref.\cite{Edwards}). 
	}
	\label{figM3}
\end{figure}
} 
\vspace{0.2in}
{
\begin{figure}
	\centering
	\includegraphics[width=0.5\textwidth]{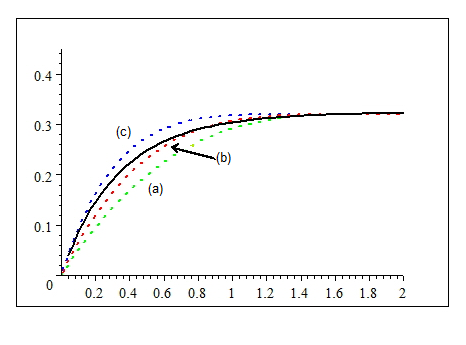}
	\vspace{-0.0in}
	\caption{\footnotesize \normalfont 
	Convenient  fit to the numerical solution by the function of eq.(\ref{form}).
	where (a) $ p=3$; (b) $p=4$;  (c) $p=6$.   We use the best fit, with $p=4$ and with $u(\infty)=0.3240$.  
	}
	\label{figsolnsfit}
\end{figure}
} 
\vspace{0.2in}

For low momentum fermions interacting
in the potential with the bound state
we have $[\bar{\psi}_L(X\!+\!r)\psi_{R}(X\!-\!r)]_{f}\approx [\bar{\psi}_L(X)\psi_{R}(X]_{f}$.
 The Yukawa coupling, using $g_0^2N/8\pi^2=f$, and setting $M_0=1$, is then given by eq.(\ref{5NJL000}):
   \bbo   
   \label{115}\backo\backo
  g_Y =  \sqrt{2N_cJ} g_0^2 \!\int\!\! d^4r D_F(2r) \phi(r) 
   =\sqrt{\frac{N_cJ}{2}} \frac{8\pi^2 f}{N_c} \!\! \int_0^\infty\!\! 4\pi \frac{e^{-2r}}{8\pi r}\phi(r) r^2 dr
   = \sqrt{\frac{N_cJ}{2}} \frac{\pi f}{N_c} \!\int_0^\infty\!\! 4\pi  \bl\frac{e^{-2r}}{r}\phi(r)\br r^2 dr.
   \ebo 
We evaluate the integrals of $\phi(r)=Au(r)/r$  over the potential, (with normalization factor $A$),
  \bbo \backo\backo
  \label{125}
    ( 4\pi A) \bl  \int_0^{R_0}\! \bl
 \frac{0.324\left(1-(1-r/R_0)^4 \right)}{ r}\br \frac{e^{-2r}}{r} r^2 dr+ \!\int_{R_0}^\infty \!
\bl\frac{0.324}{r}e^{-|\mu|(r-R_0)}\br\frac{e^{-2r}}{r} r^2 dr\br
   =  1.1412\times A.
   \ebo
   The second term is insensitive to $\mu$ for $|\mu|<\!\!< M_0$.
The normalization, as usual, is dominated by the large distance tail:
\bbo
\label{126}
   1\approx
     A^2 \bl 4\pi \int_{R_0}^\infty \bl \frac{0.324}{r}e^{-|\mu|(r-R_0)}\br^2 r^2 dr\br=0.65948  
     \frac{A^2}{|\mu|^2}
   \;\;\;\makebox{hence,}\;\;\; A=1.2313 \sqrt{\frac{|\mu|}{M_0}}, 
\ebo
(here we restore $M_0$; note that previously we defined $R_0=1.8(M_0)^{-1}$ and we
have set $M_0=1$, hence $M_0$ reemerges in the above ratio $|\mu|/M_0$).

For the Yukawa coupling,  $g_Y$ in eq.(\ref{115}), we have the prefactor, (using $N_c=3$, $J=16$, $f=1.0694$): 
   \bbo
   \label{127}
   \sqrt{\frac{N_cJ}{2}}\frac{\pi f}{N_c}=5.4862.
   \ebo
 We  thus obtain the value of $M_0$ upon combining eqs.(\ref{125}, \ref{126}, \ref{127}):
   \bbo
   g_Y = 1 = 5.4862\times 1.412 \times 1.2313 \sqrt{\frac{|\mu|}{M_0}}=7.7090\sqrt{\frac{|\mu|}{M_0}}.
   \ebo
 Upon inputting $|\mu|=88$ GeV, we obtain;
   \bbo
    M_0=5.23\;\makebox{TeV.}
   \ebo
 The ``fit'' used here was chosen for convenience and a better  fit
 can yield a more precise result.  However, this roughly confirms the typical estimates,
 $M_0 \sim 5-6$ TeV, we obtained above from the other approximations.


\vspace{0.5in}

\section{Putting it all together: \ Composite Brout-Englert-Higgs Boson and Natural Top Quark Condensation  \label{all}}

We introduced a dynamical electroweak isodoublet bilocal BEH field composed of third generation chiral quarks:
 \vspace{0.0in}    
 \bbo
M_0^2H^i(x,y) \sim [\bar\psi_R(x)\psi^i_{L}(y)],  \qquad \psi^i_L=\bl\begin{array}{c} t\\ b \end{array}\br_L,
\qquad \psi_R = t_R,
   \ebo     
[...] color contracted, $(^i)$ isospin index.
As an ansatz we follow Yukawa and assume this can be factorized \cite{Yukawa}:
\vspace{0.0in}    
 \bbo
H^i(X,r) = H^i(X)\phi(r). 
   \ebo    
 Here the primary
factor field, $H(X)$, can be viewed as the normal isodoublet BEH field and carries $SU(2)\times U(1)$ 
electroweak charges.  The internal field, 
$\phi(r)$, is complex  and, as a low energy approximation, we can assume it carries no electroweak charges.

Here  we will move the $SU(2)_L\times U(1)_Y$ left-handed gauge charges, located at $y=X+r$, and
the $U(1)_Y$ right-handed hypercharge, located at $x=X-r$, to reside at the center of the state, $X$.
This can be viewed as an approximation, but it can be formally  arranged introducing Wilson lines internal to the state,
as described in Appendix \ref{gauge}.
The configuration of the BEH boson is then a collection of
radial, L (R), Wilson lines extending respectively from $X$ to $y=X-r$ ($x=X-r$),
becoming  a ``hedge-hog''
configuration of internal Wilson lines.
This leads to the usual electroweak physics 
and gauge field mass generation of the BEH boson
with no complication from $r$ dependence of $\phi(r)$
to leading order in $1/M_0^2$.  

The present scheme
is ``minimal,'' so we suppress the $b_R$ quark in the following.
More realistically, we can assume the entire third generation experiences 
the full topcolor $SU(3)_{TC}\times U(1)'$ interaction.
Since we want to avoid a $\bar{b}_Lb_R$ condensate we typically require a
$Z'$ boson as in the old topcolor models \cite{Topcolor}\cite{Topcolor2}.  This supplies
a repulsive force in the $\bar{b}_Lb_R$ channel. 
In a more complete topcolor $SU(3)_{TC}\times U(1)'$ scheme 
we naturally have $SU(2)_L\times SU(2)_R\times U(1)$ with common $M_0$ 
for coloron and $Z'$ masses.  
To have $SU(2)_L\times U(1)$ with only a top condensate we require \cite{NSD}:
\bbo
g^2_{0} + \frac{8}{27}g^2_{U(1)'}
> g^2_{c} 
\qquad\;\;\;\;
g^2_{c} > g^2_{0} -
\frac{4}{27}g^2_{U(1)'}.
\ebo
This is easy to do without fine--tuning
and has been discussed in various model papers \cite{Topcolor}\cite{Topcolor2}\cite{NSD}.

The action for the composite BEH field then becomes,
 \bbo
\backo\backo S=
\!\!\int d^4X \bl
 |D_H H(X)|^2+ 
|H(X)|^2 \; M_0^3\!\! \int d^3r\bl
-|\partial_{\vec{r}}\phi(r)|^2 
+{g_0^2 N_c M_0}\frac{ e^{-2M_0 |\vvr|}}{8\pi |\vvr|}|\phi(\vvr) |^2\br 
\nonumbo\qquad\qquad
-\frac{\lambda}{2}(H^\dagger H)^2 - g_Y\left([\bar{\psi}_L(X)t_{R}(X)]_{f}H(X)  +h.c.\right)\br,
   \ebo     
 where the induced BEH-Yukawa  coupling, $g_Y=g_{top}\approx 1.0$, is:
\bbo 
g_Y\approx  g_0^2 \sqrt{2N_c/J}\;\phi(0),    
 \ebo
 The internal wave-function $\phi(\vec{r})$  satisfies the 
SKG equation (we can neglect small $\lambda$ corrections here)
with eigenvalue $\mu^2$:
\bbo   
\backo\!\!\!\!
-\bl\frac{\partial^2 }{\partial r^2}+\frac{2}{r}\frac{\partial }{\partial r}\br\phi(r)-g_0^2N_c M\frac{ e^{-2M_0 |\vvr|}}{8\pi |\vvr|}\phi(r) =\mu^2\phi(r).
\ebo 
This yields a compact solution, $\phi(\vec{r})$, and 
the eigenvalue $\mu^2$ is then the physical (mass)$^2$  of the bound state.
 
 For supercritical coupling, $g_0^2>g_c^2$ we have negative $\mu^2$ and
 we then obtain  the SM ``sombrero potential'' for the symmetric phase:
  \bbo
 V(H) = -|\mu|^2|H|^2+\frac{\lambda}{2}|H|^4 .
    \ebo
    Near critical coupling $\phi(r)\sim e^{-|\mu| r}/r$ with $|\mu| <\!\!<M_0$.
    We see a chiral instability and 
the field $H$  develops a VEV. 
We then conclude $v =|\mu|/\sqrt{\lambda}$, which is true for
the pointlike BEH boson.

 The quartic interaction is determined at loop level and  can be obtained in
 the full bilocal theory. We find the logarithms  match with the effective 
 pointlike theory calculation, which is simpler. 
 The Feynman  loop with four vertices integrates the loop momenta
from
$\mu^2$ to $M_0^2$ as IR and UV cut-offs (we could equally well include
$\mu^2$ in the propagator denominator for the IR cut-off  with similar results).
In the pointlike limit for the potential,
we obtain:  
\bbo
\label{quartic1}
\backo
\frac{\lambda}{2} 
= \frac{N_c}{8\pi^2 }\hat{g}_Y^4|\phi(0)|^4 
\bl\ln \left( \frac{M_0}{\mu }\right)+{\cal{O}}\left(\frac{\mu^2}{M_0^2}\right) \br  \approx \frac{N_{c}g_Y^{4}}{8\pi ^{2}}%
 \ln \left(\frac{M_0}{\mu }\right).
\ebo
The log evolution matches the result for the pointlike NJL case, with $g_{top}=g_Y=\hat{g}_Y\phi(0)$.
Not surprisingly when we take the pointlike potential limit the loop
result of the bilocal theory confirms a pointlike NJL loop calculation  with $g_Y\sim \phi(0)$ 
(see discussion of Feynman loops in Section 4 of ref.(\cite{main})). 

In the standard model the quartic coupling term in the Higgs potential is 
$\frac{\lambda}{2}(H^\dagger H)^2$.
Experimentally, in the SM using the value of $m_{BEH}\approx 125$ GeV and $v_{weak}\approx 175$ GeV, 
we find $\lambda \approx 0.25$.
In  the old pointlike NJL top condensation model the quartic coupling
was determined by the RG with ``compositeness boundary conditions,''
where we obtained (running-down from the Landau pole at $M_0$ to $|\mu|$), the result $\lambda\sim 1$.
This is too large and leads to predicted $m_{BEH}\sim 260$ GeV. Indeed, the quartic coupling is generally problematic for all 
pointlike NJL based theories of a composite BEH boson, e.g.  \cite{HMTT}.
However, in the present bilocal scheme, owing to suppression of $g_Y\sim \phi(0)$, the quartic coupling
is also suppressed $\propto |\phi(0)|^4$ and is now generated in running, from a value
of zero at $M_0$, down  to $|\mu|\sim 88$ GeV, using the induced
$g_Y$.  Keeping only the $g_Y^4$ contribution
we now obtain numerically from eq.(\ref{quartic1}): $\lambda \approx 0.32$.

This result is significantly better than the old NJL model,
but we can do better. The prefactor at one loop level should reflect the full RG running of $\lambda$,
(see, e.g., \cite{HillTA}), yielding;
\bbo
\lambda \approx (g_Y^{4}-g_Y^2\lambda-\lambda^2) \frac{N_{c}}{4\pi ^{2}}%
 \ln \left( \frac{M_0}{\mu }\right) \approx 0.23\;\;\; \makebox{(cf., $0.25 $ experiment.)};
\ebo  
This is in very good agreement with experiment at one loop precision.
It also represents a ``break-through'' in these kinds of models where, as mentioned above, it
has generally been problematic to reduce $\lambda$ much below unity when only
RG running is deployed over a large range of scale.

The key result obtained for $M_0$ comes from the numerical integration of
the SKG equation in Section \ref{numerical}, and implementing the relation between $g_Y$ and $\sqrt{|\mu|/M_0}$
with $|\mu|=88$ GeV, the symmetric phase BEH mass.  This yields $M_0= 5.23$ TeV.
 Comparable results of $\sim 5.0-6.0$ TeV are also obtained in the simple skeletal model of
 Section \ref{skeletal}, as well as for the matched spherical potential well of Section \ref{potwell}
 (in the latter case we may have  analytical control over the short-distance limit of the solution).
 We emphasize that, while we are confident of the prediction $M_0 \sim 5 - 6$ TeV,
 these results come from the present semiclassical analysis.  Quantum
 corrections will be explored elsewhere \cite{CTHprep} and may be significant.
 Due to the linear relationship between $g^2$ and $|\mu|/M_0$, a consequence
 of the dilution effect of $\phi(0)$, 
we see the degree of fine-tuning of the hierarchy is of order
$\delta \kappa/\kappa \sim |\mu|/M_0 \sim 1.4\%$ (Section  \ref{finetune}).

\newpage

\section{ Conclusions}

\noindent
A summary of the results of the bilocal theory of the BEH boson is as follows:
    
\begin{itemize}  
\item Symmetric phase BEH mass  $\mu^2 = -(88)^2$ (GeV)$^2$ is an input.
\item The top-quark-BEH-Yukawa coupling $g_Y \approx 1.0$ is an input.
\item Our approximations of bound state solutions in the Yukawa potential imply $M_0\approx 6$  TeV.
\item The fine--tuning near critical coupling is about  $\sim 1.4 \%$, significantly aided by dilution.
\item The theory ``predicts'' $\lambda \approx  0.23 $ very close to experiment $\approx 0.25$.
\item The BEH mass $\sim 125$ GeV and  weak scale $v_{weak}=175$ GeV  are then obtained a usual.
\item The main prediction for the future is an  octet of colorons, at a mass scale of order $M_0\sim 6$ TeV.

\end{itemize}

The Nambu--Jona-Lasinio model (NJL) is a
Lorentz invariant description of a scalar bound 
state of relativistic chiral fermions in 
a 4-fermion short-distance potential: $-\frac{g^2}{M_0^2}\bar\psi_L \psi_R\bar\psi_R \psi_L$.
The solution of the NJL model is constrained to be
a pointlike effective field theory, $\Phi(x)\sim \bar\psi_R(x) \psi_L(x)$, with renormalization group (RG)
boundary conditions on its parameters at $M_0$.
At critical coupling, $g_0^2\rightarrow g_c^2 = 8\pi^2/N_c$, the bound state mass $\mu^2\rightarrow 0$.
The low energy effective action then approaches a conformal theory.
Indeed, the RG parameters, top Yukawa coupling and quartic coupling,  approach IR fixed points,
(in analogy to a 2nd order phase transition at critical temperature).

The original top quark condensation (composite BEH boson) model was formulated using the NJL model
with third generation constituents \cite{Nambu1}\cite{Yama}\cite{BHL}. The model gave precise predictions,
but the values obtained for $m_{top}$ and $m_{BEH}$ were not in
agreement with experiment. Moreover, the model required an absurd degree
of fine-tuning. The authors at the time thought these issues may be resolved in future
developments of the theory. 

 In quantum mechanics a short distance potential 
 (e.g, $\sim -\alpha \delta^3(r)$), 
with eigenvalue near zero (critical coupling), will  always produce a $\sim 1/r$ large distance ``tail'' wave-function.  This is 
due to scale symmetry outside the potential.
The  NJL model is constrained to be pointlike and has no internal wave-function, therefore no IR ``tail.'' 
To obtain an internal wave-function we must extend the pointlike NJL field description 
of the bound state, $H(x)$,
to a bilocal description (similar to that suggested long ago by \cite{Yukawa})  
 $H(x)\rightarrow H(x,y)\sim \bar\psi_R(x) \psi_L(y)$. 
We write a factorized ansatz, $H(x,y)\rightarrow H(X,r)\sim H(X) \phi(r)$,
 where $X=(x+y)/2$ and $r=(x-y)/2$, and $\phi(r)$ becomes the internal wave-function
 of the bound state. 
 
 The electroweak gauging by the covariant derivatives can be assumed 
 to act upon  $H(X)$, while $\phi(r)$
 is then sterile. Formally, we can arrange this by field redefinitions using Wilson lines
 to ``pull-back'' the gauge field currents to $X$. In any case,
 this affects only the electroweak gauging at large distances and can be viewed
 as a low energy approximation.  It is sufficient to allow the composite BEH boson to develop
 a VEV in the usual way and supply masses to $W$ and $Z$ gauge bosons of the standard model. 
 
The pointlike NJL interaction is then replaced by a suitable UV completion.
Most natural is ``topcolor'' \cite{Topcolor},
consisting of a massive octet of ``colorons,'' leading to a single particle exchange potential.
The free bilocal fields must be normalized 
 to have well defined currents and charges. This is nontrivial, and
 requires removing ``relative time,'' after which $\phi(\vvr)$ becomes
 a static field with no dependence upon $r^0$.
This yields, in the rest-frame, a Yukawa potential interaction,
$ - g^2 N_cM_0 (\exp(-2M_0r)/8\pi r)|\phi(\vvr)|^2 $, with a
BCS-like enhancement $\propto N_c$ due to color singlet normalization.

The internal wave-function of the bound state then satisfies a 
semiclassical Schr\"odinger-Klein-Gordon (SKG) equation
with eigenvalue $\mu^2$ in the rest frame:
\bbo   
\backo\!\!\!\!
-\bl\frac{\partial^2 }{\partial r^2}+\frac{2}{r}\frac{\partial }{\partial r}\br\phi(r)-g_0^2N_c M\frac{ e^{-2M_0 |\vvr|}}{8\pi |\vvr|}\phi(r) =\mu^2\phi(r)
\ebo 
This yields a compact solution, $\phi(\vec{r})$, and 
the eigenvalue $\mu^2$ is then the physical (mass)$^2$  of the bound state.
The critical coupling, $g_0^2= g_c^2$, is numerically almost identical to the NJL critical coupling.
At critical coupling, $\mu^2 =0$, while at super-critical coupling, $\mu^2 < 0 $, 
implying spontaneous symmetry breaking.
Near critical coupling $\phi(r)\sim e^{-|\mu| r}/r$ with $|\mu| <\!\!<M_0$.
The solution can be reliably obtained by numerically integrating the SKG equation.

The Yukawa coupling of the bound state to free fermions is also generated by the coloron interaction,
  $g_Y \propto \phi(0)$.
Due to the extended wave-function tail, $\phi(0)\sim \sqrt{|\mu|/M_0}$, we have significant
``dilution,'' and $g_Y $ is suppressed.
The numerical integration of the solution of the SKG equation in the Yukawa potential (Section \ref{numerical}), reveals the critical behavior and  yields the result, $g_Y\propto \sqrt{|\mu|/M_0}$, and 
the value of the coloron mass/composite scale, $M_0=5.23$ TeV .
 This confirms various estimates we performed
in the $\delta-$function limit of the potential, as well as an exactly solvable spherical well approximation,
yielding of order $M_0=5.0-6.0$ TeV.

The BEH boson quartic coupling arises from loops $\sim N_c g_Y^4\ln(M/\mu)/4\pi^2 \propto (\phi(0))^4$
(Feynman loops are studied in greater detail in \cite{main}).
In application to top quark condensation, the dilution effect suppresses $g_Y$ and,
inputting  $g_Y=g_{top} = 1$, from which we determine the quartic coupling.
The result for the quartic coupling at one loop is $\lambda\sim 0.23$, compared
to $\lambda\approx 0.25$ experimentally, in excellent agreement.

Remarkably,  fine--tuning of the hierarchy, $M_0/|\mu|>\!\!> 1$, is also suppressed by dilution, 
of order $\mu/M_0 \sim (100\;GeV)/(6\; TeV) \sim$ (few) $\% $, due to an
emergent linear relation, 
$\delta g_c^2/g_c^2 \sim \mu/M_0$. This is shown to be general by the simple
spherical potential model.

The ``colorons,'' which mediate the binding interaction, form a 
QCD color octet.
They have a global $SU(3)$ symmetry and a conserved $SU(3)$ current, and must 
therefore be pair-produced.  They will decay in the minimal model to $\bar{t}t$,
but in more realistic topcolor models 
they can also decay to $\bar{b}{b}$ (\cite{Topcolor}\cite{Topcolor2}\cite{NSD} and references therein).
They may be accessible
 to the LHC in the multi-TeV range \cite{Burdman}, favoring the third generation
in its couplings.  It should be easy to obtain a lower bound on $M_0$ from single gluon or gluon fusion
 producting pairs of the colorons.  
 The colorons will produce excesses in 4-top, $t\bar{t}t\bar{t}$, events.  In
 an extended model we would also expect $t\bar{t}b\bar{b}$ and $b\bar{b}b\bar{b}$ anomalies
 to emerge.   The more general topcolor models offer many possibilities
 within the arena of heavy quark flavor physcs. 

This theory, if confirmed, solves the ``naturalness problem'' of the BEH boson in the Standard Model.
Many avenues for further theoretical development exist.  Notably, a revisitation 
of the Topcolor $Z'$  and a possible second heavy boson (resonance?)
associated with $\bar{b}b$ would be interesting to study in the present formalism. 
It may also be useful to apply these techniques in QCD, such as heavy-light meson 
and heavy-heavy-light baryon theory \cite{bardeenhill}.

\vspace{0.25in}
\appendix

\section{Gauging and Wilson Lines \label{gauge}}

We have simplified the electroweak gauging in the above definition of the BEH model,
where the covariant derivatives act only upon the center coordinate $X$ in  $H(X)$, 
as opposed to $x$ and $y$ in  $H(x,y)$.  
We can formally arrange this by incorporating ``Wilson lines'' into the
structure of the wave-function. 

Consider the $SU(2)_L$ and weak hypercharge $U(1)_Y$ covariant derivatives
(we will omit displaying the gluons in the analysis, which  ultimately cancel
at the $X$ endpoint), where $H \sim \bar{t}_R\psi_L$,
\bbo
\backo
D^\dagger_RH(x,y)=\left( \frac{\partial}{\partial x^\mu} +  ig_1 B_\mu(x) \frac{Y_R}{2}\right)H(x,y), 
\;\;\;
D_LH(x,y)=\left( \frac{\partial}{\partial y^\mu} -  ig_1 B_\mu \frac{Y_L}{2}- ig_2W^A(y)_\mu \frac{\tau^A}{2}\right)H(x,y),
\ebo
where the weak hypercharges are  $[Y,\psi_L]= Y_L\psi_L=(1/3)\psi_L, \;\;[Y, \psi_R]= Y_R\psi_R =(4/3)\psi_R,\;\; [Y,H]  =[(Y_L-Y_R ),H] = Y_HH= (-1)H$.
Note that, since $H \sim \bar{t}_R\psi_L$  the action of the derivatives on $H(x,y)$ is $D_L H$, which acts upon $\psi_L$, and $D_R^\dagger H$, which acts upon $\bar{t}_R$.

We introduce Wilson lines, $H(x,y)\rightarrow W^\dagger_R(X,x)W_L(X,y)H(x,y)$
where,
\bbo
W_L(X,y) = P\exp\bl -ig_1\frac{Y_L}{2}\int_y^X \!\! \!\! B_{\nu}(\rho) d\rho^\nu  
- ig_2\int_y^X\!\!\!\!  W^A_\nu(\rho) \frac{\tau^A}{2} d\rho^\nu \br
\nonumbo
W^\dagger_R(X,x) = P\exp\bl +ig_1\frac{Y_R}{2}\int_x^X \!\! \!\! B_{\nu}(\rho) d\rho^\nu \br,
\ebo
and $P$ denotes ``path ordering'' (which is relevant for the non-abelian components, but
trivial for $U(1)_Y$).
The derivative terms become, with $X=(x+y)/2$ and $r=(x-y)/2$
and noting sign flip of the gauge fields in $D_R^\dagger$:
\bbo
\backo\backo\;\;\;
D^\dagger_R(W^\dagger_RW_LH(x,y))=
W^\dagger_RW_L\left(\frac{\partial}{\partial x^\mu} 
+ig_1\frac{Y_R}{2}\frac{\partial X_\mu}{\partial x_\nu } B_{R\nu}(X)
-ig_1\frac{Y_L}{2}\frac{\partial X_\mu}{\partial x_\nu} B_{R\nu}(X)
-ig_2\frac{\partial X_\mu}{\partial x_\nu} W^A_\nu (X)\frac{\tau^A}{2}
\right)H(x,y)
\nonumbo
=W^\dagger_RW_L\left(\frac{\partial}{\partial x^\mu} 
+i\half  g_1 \bl\frac{Y_R}{2} - \frac{Y_L}{2}\br  B_\mu(X)- i\half g_2 W^A_\nu(X) \frac{\tau^A}{2}\right)H(x,y)
\nonumbo
=W^\dagger_RW_L\left(\frac{\partial}{\partial x^\mu} 
-i\half  g_1\frac{Y_H}{2}  B_\mu(X)- i\half g_2 W^A_\nu(X) \frac{\tau^A}{2}\right)H(x,y),
\ebo
(note, e.g., $\partial_{x^\mu}(\int_x^X \! B_{\nu}(\rho) d\rho^\nu)=\half B_\mu(X)-B_\mu(x)$, etc.). 
Likewise we obtain:
\bbo
D_LH
=W^\dagger_RW_L\left(\frac{\partial}{\partial y^\mu} 
-i\half g_1 \frac{Y_H}{2}  B_\mu(X) - i\half g_2 W^A_\nu(X) \frac{\tau^A}{2} \right)H(x,y),
\ebo
Pass to barycentric coordinates and $H'(X,r))=H(X+r, X-r)$,
\bbo
D^\dagger_R(W^\dagger_R W_L)H'=
 \half W^\dagger_RW_L  \bl  \frac{\partial}{\partial X^\mu}+\frac{\partial}{\partial r^\mu} 
- ig_1 \frac{Y_H}{2}   B_\mu(X) -i g_2 W^A_\nu(X) \frac{\tau^A}{2}\br H'(X,r),
\nonumbo 
D_L(W^\dagger_R W_L)H'=
 \half W^\dagger_RW_L  \bl  \frac{\partial}{\partial X^\mu}-  \frac{\partial}{\partial r^\mu} 
- ig_1 \frac{Y_H}{2}   B_\mu(X) -i g_2 W^A_\nu(X) \frac{\tau^A}{2}\br H'(X,r).
\ebo
Note that each term has acquired the overall factor of $\half$.
We assume factorization, $H(x,y)\rightarrow \sqrt{2/J}\;H(X) \phi(r)$ and
the overall kinetic term action becomes:
\bbo
\int d^4X d^4r\; \left| \partial_X H(X) -  ig_1 \frac{Y_H}{2}  B_\mu(X)H(X)- ig_2 W^A_\nu(X) \frac{\tau^A}{2} H(X)\right|^2
|\phi(r)  |^2
+|\partial_r \phi|^2|\chi|^2\br,
\ebo
showing that the Wilson lines ``pull back'' the gauge fields
to the center $X$ and assemble the component hypercharges into the BEH bound state $Y_H=Y_L-Y_R$.
We have omitted consideration of the QCD terms since they vanish for the same reason
that they cancel in a local expression, $\partial_x (\bar\psi_R(x)\psi_L(x))$.

This redefines the $H$
field with the Wilson lines, 
and recovers the gauging we proposed above for our composite BEH theory.
The symmetry breaking and masses of the gauge fields go through in the usual way.

\section{Lorentz Invariance of $\mu^2$ and Covariantization \label{LI}}

We show that the resulting eigenvalue $\mu^2$ is Lorentz invariant, resulting from the original 
manifestly Lorentz invariant factorized action of 
 eq.(\ref{af2}),
 \bbo
\label{C1}
 \backo S= M^4
\!\!\int\!\! d^4X d^4r\; 
\bl Z|\phi(r)|^2|\partial_X\chi(X)|^2 
+Z|\chi(X)|^2|\partial_r\phi(r)|^2 
+2{g_0^2 N_c}D_F(2r)|\chi(X)\phi(r) |^2 
\br.
\ebo   
First, we see that
$\chi$  can be defined to have canonically normalized $\chi(X)$
 kinetic term by introducing a Lagrange multiplier, as:
\bbo
\label{eqL}\backo\backo
S_1= \eta_1 
\bl 1- M^4\!\!\int\! d^4r \;Z|\phi(r)|^2  \br^2,
\ebo
 and $\delta S_1/\delta \eta_1 = 0$.
We  define currents (these are discussed in Appendix B of ref.\cite{main}):
\bbo
J_\mu = i[\chi^\dagger(X) 
 \frac{ \stackrel{\leftrightarrow}{\partial} }{ \partial X^\mu }\chi(X)],
 \qquad \qquad
 K_\mu =i[\phi^\dagger(r) 
 \frac{\stackrel{\leftrightarrow}{\partial}}{\partial {r^\mu}}\phi(r)],
\ebo
and we can therefore define a timelike unit vector, $\omega_\mu$:\footnote{The bilocal currents are,
  \bbo
 J^+_{\mu}(X,r)
 =iZ'\epsilon^4 M^4[\chi^\dagger(X) 
 \frac{ \stackrel{\leftrightarrow}{\partial} }{ \partial X^\mu }\chi(X)]\phi^\dagger(r)\phi(r),
 \qquad
 J^-_{\mu}(X,r)
 =iZ'M^4[\phi^\dagger(r) 
 \frac{\stackrel{\leftrightarrow}{\partial}}{\partial {r^\mu}}\phi(r)]\chi^\dagger(X)\chi(X)
 \nonumber \ebo 
 These can be integrated to form, 
 \bbo
 J^+_{\mu}(X)
 =iZM^4[\chi^\dagger(X) 
 \frac{ \stackrel{\leftrightarrow}{\partial} }{ \partial X^\mu }\chi(X)]\!\int\!\! d^4r\; \phi^\dagger(r)\phi(r),
 \qquad
 J^-_{\mu}(r)
 =iZM^4[\phi^\dagger(r) 
 \frac{\stackrel{\leftrightarrow}{\partial}}{\partial {r^\mu}}\phi(r)]\!\int\!\! d^4X \;\chi^\dagger(X)\chi(X)
 \nonumber \ebo 
 Normalizing
 \bbo
 1=ZM^4\int\!\! d^4r\;|\phi(r)|^2=M^3\int\!\! d^3r|\phi(r)|^2
 \qquad 
 \makebox{hence, } \qquad
 J_{\mu}(X)
 ={i}\chi^\dagger(X) 
 \frac{ \stackrel{\leftrightarrow}{\partial} }{ \partial X^\mu }\chi(X).
 \nonumber \ebo  }
\bbo
0= \omega _\mu\sqrt{J_\rho J^{\rho}} - J_\mu.
\ebo
We can formally implement the constraint that $\phi(r)$ has no
dependence upon $r^0$ by adding
to the action a Lagrange multiplier, $\eta$,  while
preserving Lorentz invariance,
\bbo
\label{eqL}\backo\backo
S_2= \eta_2 \!\!\int\!\!d^4X d^4r\; M^4 |\omega^\mu K_\mu|^2,\;\;
\ebo\normalsize
 and $\delta S_\eta/\delta \eta = 0$.
 The constraint implies that $\omega^\mu \partial_\mu \phi(r)=0$ where $\omega^\mu \propto P^\mu$
 is the timelike 4-momentum of the bound state, hence $\phi$ has no $r^0$
 dependence in the rest frame.  Note that,  in practice, we don't need
 this formality if we simply assume 
 that we are interested only in the solutions in which
 $\phi(r)$ has no dependence upon $r^0$.  
 
 Now consider the kinetic terms,
 \bbo
\label{}
 \backo S= M^4
\!\!\int \!\! d^4X d^4r\; 
\bl Z|\phi(r)|^2|\partial_X\chi(X)|^2 
+Z|\chi(X)|^2|\partial_r\phi(r)|^2 \br.
\ebo 
$Z$ can be interpreted as an operator of the form, 
\bbo
Z\rightarrow \delta (M_0\omega_\mu r^\mu), 
\ebo
which
 removes the relative time in the kinetic terms
in the rest-frame:
\bbo
\label{}
 \backo S\rightarrow  M^3
\!\!\int\!\! d^4X d^3r\; 
\bl |\phi(\vvr)|^2|\partial_X\chi(X)|^2 
-|\chi(X)|^2|\partial_{\vvr}\phi(\vvr)|^2 \br,
\ebo
and the prefactor is now $M^3$
 (Note $|\partial_r\phi|^2= |\partial_{r^0}^2\phi|^2 - |\partial_{\vec{r}}\phi|^2$
 where $\partial_{\vec{r}}$ is the spatial derivative).
 
 If we examine the constraint $S_1$ we see that,
 \bbo
\label{}\backo\backo
S_1= \eta_1 
\bl 1- M_0^4\!\!\int\!\! d^4r \;Z|\phi(r)|^2  \br^2\;
\longrightarrow \eta_1 \;
\bl 1- M_0^3\!\!\int\!\! d^3r\; |\phi(\vvr)|^2\br^2,
\ebo
 which enforces the rest frame normalization $ 1= M^3\!\!\int\! d^3r |\phi(\vvr)|^2$.
 The kinetic terms become:
 \bbo
\label{}
 \backo S\rightarrow  
\!\!\int\!\! d^4X  \;
|\partial_X\chi(X)|^2  
-M_0^3\!\!\int d^4X d^3r\; |\chi(X)|^2|\partial_{\vvr}\phi(\vvr)|^2. 
\ebo
With the timelike unit vector we 
can define a tensor,
\bbo
\label{tensor}
W_{\mu\nu}= \omega_{\mu}\omega_\nu -g_{\mu\nu}, 
\ebo 
hence,
\bbo
\label{B12}
\backo\backo
(r^0)^2 = (W_{\mu\nu}+g_{\mu\nu}) r^\mu r^\nu; \qquad
{\vvr}{\;}^2 \equiv W_{\mu\nu}r^\mu r^\nu; \qquad 
\phi(r)\equiv \phi(\sqrt{W_{\mu\nu}r^\mu r^\nu}); 
\qquad 
W^{\mu\nu}\partial_\mu\phi^\dagger\partial_\nu \phi = -|\partial_{\vvr}\phi|^2.
\ebo
Using $W^{\mu\nu} $, all of the expressions in eq.(\ref{C1}) can be made manifestly Lorentz invariant.

Hence, from eqs.(\ref{eqL},\ref{B12}) with $r=\sqrt{W_{\mu\nu}r^\mu r^\nu}$, we see that $\mu^2$ is given by the manifestly invariant form:
 \bbo
  \label{intDf}
\mu^2 M_0^4\!\int \!\! d^4r 
\delta (M_0\omega_\mu r^\mu) |\phi({r})|^2 =  
 M_0^4\!\!\int \!\! d^4r\; \left(
\delta (M_0\omega_\mu r^\mu)|\partial_{r}\phi(r)|^2 
+\;{2g_0^2 N_c M }D_F(2r^\mu)|\phi(r)\; |^2
\right)\br
\nonumbo
 = \mu^2
 =M_0^3\!\int \!\! d^3r \; \left( -\phi^*(\vec{r})\nabla^2_{\vec{r}}\phi(\vec{r})-
{g_0^2 N_c  M_0}\frac{ e^{-2M_0 |\vvr|}}{8\pi |\vvr|}|\phi(\vvr) |^2
\right)
 \ebo 
Once calculated in the rest frame it is the same in any frame. 


\section{The Vacuum as a Bose-Einstein Condensate: Frame Averaging \label{CV}}

We have thus far been working in the rest frame of the BEH boson
in the symmetric phase.  Here the BEH boson has nonzero timelike
4-momentum, since $P^2=\mu^2$, and $P=(\mu,\vec{0})$, 
putting aside the fact that $\mu^2<0$ and $\mu$ is imaginary.
The broken phase state has $\phi(\vvr)$ with normalizable 
wave-function and has a definite $\vvr$ dependence. 
We have upon removal of relative time, $P_\mu r^\mu =0$,
hence $\phi(\vvr)$ has a purely spacelike argument in the particle rest frame.
As the field $\phi$ acquires a VEV, where
then $P_\mu$ vanishes,  what would select the frame for $\phi(\vvr)$? 

In the SM there is no preferred reference frame for the broken phase that defines
the vacuum of the standard model.
That is, the BEH vacuum expectation value (VEV) in the SM is $v_{weak}$ and is a constant in space-time.
Our present theory would lead to a constant $v_{weak}$, but may contain a nonzero correlator, $\phi({r^\mu})$.
If the vacuum of the theory has an internal wave-function with a frame dependent $\phi(\vvr)$
it would imply 
Lorentz violation and observable features.  
Most parameters (Yukawa and quartic couplings) 
depend only upon $\phi(0)$
in the large $M_0$ limit, but perhaps 
unwanted suppressed effects of ${\cal{O}}(\mu^2/M_0^2)$ arise?
Presently we will describe a simple proposal, but this is a preliminary ``sketch'' and will
be (or an alternative will be) developed elsewhere \cite{CTHprep}. 

The vacuum  is akin to a condensed matter state, such as  a BEC (Bose-Einstein Condensate).
For any given composite bilocal BEH boson in its rest frame, defined by $\omega_\mu$,
we have 
$\chi(X)\phi(r)\rightarrow \chi(X)\phi_0(\vvr)= \chi(X)\phi_0(\hat{r}) $.
where $\hat{r} = W_{\mu\nu}\omega^\mu r^\nu$. We propose that the BEC is a coherent state 
of BEH bosons with arbitrary $\omega^\mu$.
We should therefore average the vacuum over $\omega^\mu$,
or ``frame average''
amplitudes that dpend upon 
$\phi(\hat{r})$.


We can give various definitions of frame averaging integral. For example,
we might integrate over the space-like hyperboloid defined by  $\omega^2=1$,
then define,
\bbo
\langle F(r^2)\rangle=
N\int \!\!d^4\omega\; \delta(\omega^2-1) F(W^{\mu\nu} r_\mu r_\nu) 
\;\;\;\makebox{where,}\;\;\; N^{-1} = \int d^4\omega \delta(\omega^2-1).
\ebo
Upon averaging $F(r^2)$ we will have 
$\langle F(r^2)\rangle $
as a function of the invariant $r^\mu r_\mu$.
The averaging integral is an analytic function of the metric,
$g_{\mu\nu}\sim (1,-1,-1,-1)$ which is continued as $\rightarrow -\eta_{\mu\nu}\sim -(1,1,1,1)$.
We then replace $\eta$ by $g$. Alternative averaging functions could be defined.

In the Yukawa interaction of eq.(\ref{5NJL000}) we have,
\bbo 
\label{}
\backo
S_Y' =\sqrt{2N_cJ} g^2_0  M_0^2
\!\! \int\!\! d^4X d^4r \;[\bar{\psi}_L(X\!+\!r)\psi_{R}(X\!-\!r)]_{f}D_F(2r)\;\chi(X)\phi(\vvr) {+h.c.}
\nonumbo
=\hat{g}_Y J M_0^2
\!\! \int\!\! d^4X d^4r \;[\bar{\psi}_L(X)\psi_{R}(X)-r^\mu r^\nu\frac{\partial}{\partial X^\mu}\bar{\psi}_L( X)\frac{\partial}{\partial X^\nu}\psi_{R}(X)+...]_{f}D_F(2r)\;\chi(X)\phi(\vvr) {+h.c.},
\ebo
thus we encounter terms in the expansion in the broken phase, such as,
\bbo 
\label{}
\backo
=-\hat{g}_Y M_0^2
\!\! \int\!\! d^4X d^4r \; [\frac{\partial}{\partial X^\mu}\bar{\psi}_L( X)\frac{\partial}{\partial X^\nu}\psi_{R}(X)+...]_{f}D_F(2r)
v_{weak} r^\mu r^\nu\phi(\vvr).
\ebo
These terms are suppressed since $D(2r)\rightarrow J^{-1}M_0^{-2}\delta^4(r)$
in the pointlike potential limit; but subleading $M_0^{-4}$ effects may remain.
If the vacuum was defined in a particular frame with a definite $\omega_\mu$
then these terms would lead to Lorentz violation. For example, the top quark  (or any other fermion
 such as  the electron),  propagating through the medium
with 4-momentum $p^\mu$ would acquire mass corrections $\propto \int_r p^\mu p^\nu r_\mu r_\nu \sim m_e^2\vec{p\;}^2/M_0^2$.
However, upon frame averaging the $\int_r \langle \propto p^\mu p^\nu r_\mu r_\nu \rangle 
\sim p_\mu p^\mu/M_0^2$ is Lorentz invariant.

However, an intriguing possibility arises in that  there may be imprinting of the cosmic reference frame at the time the vacuum forms in the early universe.  The covariant tensor, eq.(\ref{tensor})
may pick up a component of $\epsilon' T^c_{\mu\nu}$, the cosmic background stress tensor. 
We've briefly looked at this quantitatively, following Coleman and Glashow \cite{ColemanGlashow},
and we were initially surprised that electron vacuum Cerenkov radiation limits are satisfied.
From \cite{ColemanGlashow}
 we have for a particle of mass $m$ the modified dispersion
 relation, $ E^2-(1+\delta)\vec{p\;}^2=m^2$,
where $\delta$ parameterizes Lorentz breaking and would lead to
vacuum Cerenkov radiation if present.
If the vacuum wave-function has residual $\vec{r}$ dependence then we would have
nonzero $\delta$.  
We find, however,  that with $\epsilon'\sim 1$, that $\delta$
is sufficiently suppressed for the electron (which gets
mass from the BEH boson with $\phi(r)$ via with higher dimension
Eichten-Lane operators \cite{Eichten}).  We estimate $\delta \sim 
{m_e^2}/{M_0^2} \sim 10^{-14}$, below the limit $\sim 10^{-12}$ quoted in  \cite{ColemanGlashow}.

However, this is not the whole story since, for the top quark, we expect large $\delta$.
This may lead potentially to a large 
loop induced magnetic  $\vec{B}^2$ correction to the electromagnetic
kinetic term, the most sensitive probe as identified by Kostelecky \cite{Kostelecky}.
We have not done this calculation, but estimates appear problematic.
Hence a vacuum frame averaged condensate appears preferable, with a mechanism
for suppressing $\epsilon'\sim 0$. 


\newpage

\section{ Summary of the Top-BEH subsystem of the Standard Model \label{wallet}}

Lagrangian in Symmetric Phase:
 \bbo
 \backo
(DH)^\dagger DH - \mu^2 H^\dagger H - \frac{\lambda}{2}(H^\dagger H)^2- (g_t [\bar{\psi}_Lt_R]H + h.c.)
+ \bar{\psi}_L  \slash{D} \psi_L + \bar{t}_R\slash{ D}\psi_R  
\nonumbo \backo
 H= \bl\begin{array}{c} H^0\\ H^- \end{array}\br,
 \qquad D_\mu=\partial_\mu- ig_2W^A_\mu \frac{\tau^A}{2}-ig_1 B_\mu \frac{Y}{2},
 \qquad
 \psi^i_L= \bl\frac{1-\gamma^5}{2}\br \bl\begin{array}{c} t\\ b \end{array}\br,
 \qquad t_R=\bl\frac{1+\gamma^5}{2}\br t 
 \nonumbo
 \backo
 \mu^2\approx  -(88)^2\; GeV^2, \qquad g_t\approx 1, \qquad \lambda \approx 0.25, \qquad Q = \frac{\tau^3}{2}+\frac{Y}{2}
 \qquad \makebox{ $[\bar\psi \psi]= \bar\psi^a\psi_a$ denotes color sum. } 
\ebo

Lagrangian in Broken  Phase:
 \bbo \backo
\half (\partial h)^2 - \half m^2 h^2  - m_t [\bar{t}t]- \frac{g_t}{\sqrt{2}} [\bar{t}t]h - \frac{\lambda}{8}(h)^4...
\nonumbo\backo
 H= \bl\begin{array}{c} v_{weak} + \frac{h}{\sqrt{2}}+ i\phi^0 \\ \phi^- \end{array}\br
 \qquad \makebox{$\phi^0, \phi^\pm$ massless Nambu-Goldstone modes ``eaten'' by $Z^0$ and $W^\pm$ } 
 \nonumbo \backo
 m^2\approx (125)^2\; GeV^2 \qquad v_{weak}=\frac{\mu}{\sqrt{\lambda}}\approx 175\; GeV \qquad m_t= g_tv_{weak} \qquad  \lambda \approx 0.25.
\ebo


\section*{Acknowledgments}
I thank  Bill Bardeen, Bogdan Dobrescu and Julius Kuti for
discussions.  I also thank
the  Fermi Research Alliance, LLC under Contract No.~DE-AC02-07CH11359 
with the U.S.~Department of Energy, 
Office of Science, Office of High Energy Physics,
and The University of Wisconsin Physics Department for an Honorary Fellowship.


\begin{thebibliography}{99}

\bibitem{Nambu1} Y. Nambu, ``Bootstrap Symmetry Breaking in Electroweak Unification,''

Enrico Fermi Institute Preprint, 89-08 (1989). 

\bibitem{Yama}
V.~A.~Miransky, M.~Tanabashi and K.~Yamawaki,
Mod. Phys. Lett. A \textbf{4}, 1043 (1989);

{\em ibid}, Phys. Lett. B \textbf{221}, 177-183 (1989)


\bibitem{BHL}
  W.~A.~Bardeen, C.~T.~Hill and M.~Lindner,
  Phys.\ Rev.\  D {\bf 41} (1990) 1647.
  
\bibitem{CTH}
  W.~A.~Bardeen and C.~T.~Hill,
  Adv.\ Ser.\ Direct.\ High Energy Phys.\  {\bf 10}, 649 (1992);
  
 C. T. Hill, Mod. Phys. Lett. A, {\bf 5}, 2675-2682 (1990).


\bibitem{Topcolor}
C.~T.~Hill,
Phys. Lett. B \textbf{266}, 419-424 (1991);

{\em ibid}, 
Phys. Lett. B \textbf{345}, 483-489 (1995).

\bibitem{Topcolor2}



B.~A.~Dobrescu and C.~T.~Hill,
Phys. Rev. Lett. \textbf{81}, 2634-2637 (1998)

R.~S.~Chivukula, B.~A.~Dobrescu, H.~Georgi and C.~T.~Hill,
Phys. Rev. D \textbf{59}, 075003 (1999);

H.~J.~He, C.~T.~Hill and T.~M.~P.~Tait,
Phys. Rev. D \textbf{65}, 055006 (2002)

The full Fierz rearrangement beyond the most attractive channel is utilized here:

C.~T.~Hill, D.~C.~Kennedy, T.~Onogi and H.~L.~Yu,
Phys. Rev. D \textbf{47}, 2940-2948 (1993).

\bibitem{NSD}
C.~T.~Hill and E.~H.~Simmons,
Phys. Rept. \textbf{381}, 235-402 (2003); 

erratum: Phys. Rept. \textbf{390}, 553-554 (2004).


\bibitem{NJL} 
  Y. Nambu and G. Jona-Lasinio,
Phys. Rev.{\bf 122}, 345-358 (1961),

{\em ibid},
Phys. Rev. {\bf124}, 246-254 (1961),


\bibitem{Manohar}
A.~Manohar and H.~Georgi,
Nucl. Phys. B \textbf{234}, 189-212 (1984)


\bibitem{Bijnens}
J.~Bijnens, C.~Bruno and E.~de Rafael,
Nucl. Phys. B \textbf{390}, 501-541 (1993);

J.~Bijnens,
Phys. Rept. \textbf{265}, 369-446 (1996);

\bibitem{NJLReview}
See reviews and references in:

S.~P.~Klevansky,
Rev. Mod. Phys. \textbf{64}, 649-708 (1992);

U.~Vogl and W.~Weise,
Prog. Part. Nucl. Phys. \textbf{27}, 195-272 (1991);

M.~Buballa,
Phys. Rept. \textbf{407}, 205-376 (2005).

\bibitem{bardeenhill}
W.~A.~Bardeen and C.~T.~Hill,
Phys. Rev. D \textbf{49}, 409-425 (1994);

W.~A.~Bardeen, E.~J.~Eichten and C.~T.~Hill,
Phys. Rev. D \textbf{68}, 054024 (2003).


\bibitem{Schrodinger}
E.~Schr\"odinger,
Naturwiss. \textbf{14}, 664-666 (1926)
doi:10.1007/BF01507634



\bibitem{main}
C.~T.~Hill,
Nucl. Phys. B \textbf{1011}, 116788 (2025) (Steven Weinberg Memorial Volume).


\bibitem{scalars}
C.~T.~Hill,
Entropy \textbf{26}, no.2, 146 (2024).
[arXiv:2310.14750 [hep-ph]].

{\em ibid},
``Nambu and Compositeness,''
 arXiv:2401.08716 [hep-ph]].

\bibitem{Weinberg}
S. Weinberg, Quantum Theory of Fields, Vol. 1,  Cambridge University press, 1995, (pages 560, 564);



\bibitem{Feynman}
R.~P.~Feynman, M.~Kislinger and F.~Ravndal,
Phys. Rev. D \textbf{3}, 2706-2732 (1971);

\bibitem{BagModels}
R.~Van Royen and V.~F.~Weisskopf,
Nuovo Cim. A \textbf{50}, 617-645 (1967)
[erratum: Nuovo Cim. A \textbf{51}, 583 (1967)];

A.~Chodos, R.~L.~Jaffe, K.~Johnson, C.~B.~Thorn and V.~F.~Weisskopf,
Phys. Rev. D \textbf{9}, 3471-3495 (1974);

J.~Kuti and V.~F.~Weisskopf,
Phys. Rev. D \textbf{4}, 3418-3439 (1971);

W.~A.~Bardeen, M.~S.~Chanowitz, S.~D.~Drell, M.~Weinstein and T.~M.~Yan,
Phys. Rev. D \textbf{11}, 1094 (1975);

\bibitem{CJT} 
J.~M.~Cornwall, R.~Jackiw and E.~Tomboulis,
Phys. Rev. D \textbf{10}, 2428-2445 (1974).



\bibitem{Simmons}
For ``coloron phenomenology'' see:
E.~H.~Simmons,
Phys. Rev. D \textbf{55}, 1678-1683 (1997);

R.~Sekhar Chivukula, P.~Ittisamai and E.~H.~Simmons,
Phys. Rev. D \textbf{91}, no.5, 055021 (2015);

Y.~Bai and B.~A.~Dobrescu,
JHEP \textbf{04}, 114 (2018).



\bibitem{Yukawa}
H.~Yukawa,
Phys. Rev. \textbf{77}, 219-226 (1950);

{\em ibid}, 
Phys. Rev. \textbf{80}, 1047-1052 (1950);
Phys. Rev. \textbf{91}, 415 (1953);

Yukawa introduced an imaginary relative time which we abandon
in favor of relative time constraints.

\bibitem{BCS} J.~Bardeen, L.~N.~Cooper and J.~R.~Schrieffer,
Phys. Rev. \textbf{108}, 1175-1204 (1957),

L.~N.~Cooper,
Phys. Rev. \textbf{104}, 1189-1190 (1956).


\bibitem{CTHprep} C.T. Hill in preparation.

\bibitem{Dirac}
For extensive discussion of the relative time problem see:
 
K. Kamimura, 
Prog. Phys. 58 (1977) 1947;

D. Dominici, J. Gomis and G. Longhi, Nuovo Cimento B48 (1978)257; 

{\em ibid}, 
1, Nuovo Cimento B48(1978)152-166; 

 A.~Kihlberg, R.~Marnelius and N.~Mukunda,
Phys. Rev. D \textbf{23}, 2201 (1981);

H.~W.~Crater and P.~Van Alstine,
Annals Phys. \textbf{148}, 57-94 (1983);

H.~Crater, B.~Liu and P.~Van Alstine,
[arXiv:hep-ph/0306291 [hep-ph]]; 

These are usually implemented in Hamiltonian Dirac constraint theory:

P.A.M. Dirac, Can. J. Math. 2, 129 (1950); 

{\em ibid} ,`` Lectures on Quantum Mechanics'' 
(Yeshiva University, New York, 1964).

Where necessary, we prefer to implement constaints in an action formalism.


\bibitem{Wilson}
K. G. Wilson,
Phys. Rev. D {\bf 3}, 1818 (1971)

K.~G.~Wilson and J.~B.~Kogut,
Phys. Rept. \textbf{12}, 75-199 (1974).


\bibitem{PR} 
  B.~Pendleton and G.~G.~Ross,
  Phys.\ Lett.\  {\bf 98B}, 291 (1981). 
  
  C.~T.~Hill,
  Phys.\ Rev.\ D {\bf 24}, 691 (1981).

\bibitem{Eichten}
S.~Dimopoulos and L.~Susskind,
Nucl. Phys. B \textbf{155}, 237-252 (1979);

E.~Eichten and K.~D.~Lane,
Phys. Lett. B \textbf{90}, 125-130 (1980);

{\em ibid}, 
E.~Eichten and K.~D.~Lane,
Phys. Lett. B \textbf{90}, 125-130 (1980).




\bibitem{hasenfratz}
A.~Hasenfratz, P.~Hasenfratz, K.~Jansen, J.~Kuti and Y.~Shen,
Nucl. Phys. B \textbf{365}, 79-97 (1991)

\bibitem{HillTA}
C.~T.~Hill,
Phys. Rev. D \textbf{89}, no.7, 073003 (2014)



\bibitem{Edwards}  
J. P. Edwards, {\em et al.}, ``The Yukawa potential: 
ground state energy and critical screening,''\\
Progress of Theoretical and Experimental Physics, 
Volume 2017, Issue 8, August (2017), 
083A01;\\(Edwards  uses the clever curve flattening method  
for $u(r)=r\phi(r)$ to
obtain the best value of the Yukawa \\ critical screening length which
we convert to $g_c$; we do not however agree with his plots of $u(r)$.)
\\
M.~Napsuciale and S.~Rodriguez,
Phys. Lett. B \textbf{816}, 136218 (2021)
[arXiv:2012.12969 [hep-ph]].


\bibitem{HMTT}
  C.~T.~Hill, P.~A.~N.~Machado, A.~E.~Thomsen and J.~Turner,
  
  Phys.\ Rev.\ D {\bf 100}, no. 1, 015015 (2019).
  
{\em ibid.,,}
  Phys.\ Rev.\ D {\bf 100}, no. 1, 015051 (2019).
  
  C.~T.~Hill,
  ``The Next BEH Boson(s) and a BEH-Yukawa Universality,''
  arXiv:1911.10223 [hep-ph].

\bibitem{Burdman}

S.~Dawson,
Nucl. Phys. B \textbf{359}, 283-300 (1991);


C.~T.~Hill and S.~J.~Parke,
Phys. Rev. D \textbf{49}, 4454-4462 (1994)

T.~M.~P.~Tait and C.~P.~Yuan,
Phys. Rev. D \textbf{63}, 014018 (2000)

H.~J.~He, N.~Polonsky and S.~f.~Su,
Phys. Rev. D \textbf{64}, 053004 (2001)

M.~Carena and H.~E.~Haber,
Prog. Part. Nucl. Phys. \textbf{50}, 63-152 (2003);

D.~Alves \textit{et al.} [LHC New Physics Working Group],
J. Phys. G \textbf{39}, 105005 (2012)

B.~Bellazzini, C.~Csaki, J.~Hubisz, J.~Serra and J.~Terning,
JHEP \textbf{11}, 003 (2012);

G.~Isidori, A.~V.~Manohar and M.~Trott,
Phys. Lett. B \textbf{728}, 131-135 (2014);

S.~Dawson, S.~Homiller and S.~D.~Lane,
Phys. Rev. D \textbf{102}, no.5, 055012 (2020);

A.~Banerjee, S.~Dasgupta and T.~S.~Ray,
Phys. Rev. D \textbf{104}, no.9, 095021 (2021);

P.~Bittar and G.~Burdman,
[arXiv:2204.07094 [hep-ph]].

\bibitem{ColemanGlashow}
S.~R.~Coleman and S.~L.~Glashow,
Phys. Rev. D \textbf{59}, 116008 (1999).

\bibitem{Kostelecky}
V.~A.~Kostelecky,
Phys. Rev. D \textbf{69}, 105009 (2004).



\end{thebibliography}
\end{document}